\documentclass[pra,aps,twocolumn,superscriptaddress]{revtex4-2}
\usepackage{hyperref}
\usepackage{bm}
\usepackage{epsfig}
\usepackage{amssymb}
\usepackage{graphicx}
\usepackage{mathrsfs}
\usepackage{amsmath}
\usepackage{marvosym}
\usepackage{bbm}
\usepackage[mathscr]{eucal}

\begin{document}

\title{Convicting emergent multipartite entanglement with evidence from a  partially blind witness}
\author{Viktor Nordgren}
\affiliation{School of Physics and Astronomy, University of St.
Andrews, North Haugh, St. Andrews, Fife, KY16 9SS, Scotland}
\author{Olga Leskovjanov\'{a}}
\affiliation{Department of Optics, Palack\' y University, 17.
listopadu 12,  779~00 Olomouc, Czech Republic}
\author{Jan Provazn\'{\i}k}
\affiliation{Department of Optics, Palack\' y University, 17.
listopadu 12,  779~00 Olomouc, Czech Republic}
\author{Natalia Korolkova}
\affiliation{School of Physics and Astronomy, University of St.
Andrews, North Haugh, St. Andrews, Fife, KY16 9SS, Scotland}
\author{Ladislav Mi\v{s}ta, Jr.}
\affiliation{Department of Optics, Palack\' y University, 17.
listopadu 12,  779~00 Olomouc, Czech Republic}

\date{March 12, 2021}

\begin{abstract}
Genuine multipartite entanglement underlies correlation experiments
corroborating quantum mechanics and it is an expedient empowering many quantum
technologies. One of many counterintuitive facets of genuine multipartite
entanglement is its ability to exhibit an emergent character, that is, one can
infer its presence in some multipartite states merely from a set of its
separable marginals. Here, we show that the effect can be found also in the
context of Gaussian states of bosonic systems. Specifically, we construct
examples of multimode Gaussian states carrying genuine multipartite entanglement
which can be verified solely from separable nearest-neighbour two-mode
marginals. The key tool of our construction is a genuine multipartite
entanglement witness acting only on some two-mode reductions of the global
covariance matrix, which we find by a numerical solution of a semi-definite
programme. We also propose an experimental scheme for preparation of the
simplest three-mode state, which requires interference of three correlatively
displaced squeezed beams on two beam splitters. Besides revealing the concept of
emergent genuine multipartite entanglement in the Gaussian scenario and bringing
it closer to experimentally testable form, our results pave the way to effective
diagnostics methods of global properties of multipartite states without complete
tomography.

\end{abstract}
\maketitle

\section{INTRODUCTION}\label{sec_introduction}

Like from an incomplete puzzle, we assemble reality from fragments of information incoming to us from the outside world.
This coarse-grained grasping of reality is mostly sufficient for successful and safe orientation in our
environment. Barring wrong interpretation of reality, the exception to this rule may occur in situations
when the partial information available to us carries no signatures of a global property, the knowledge of which
is crucial for our correct decision.

There is a parallel with quantum world here. Namely, the wave function contains all available information about
a state of a quantum system, but for many tasks we do not need to know it completely. However, unlike in the classical world,
the ``fragments'' of the wave function may not carry traces of the global property which is important for the particular
task, yet our knowledge gained from its parts can still be sufficient.

The states with that remarkable property share similarity with entangled states \cite{Schrodinger_35} as both exhibit a counterintuitive
relationship between the whole and its parts. It is not surprising then, that examples of states with a global property which can be inferred
from parts lacking the property were nonlocal correlations \cite{Wurflinger_12,Vertesi_14} and in particular multipartite entanglement
\cite{Toth_05,Toth_07,Toth_09,Chen_14,Miklin_16,Paraschiv_17}. Out of the many flavours of multipartite entanglement \cite{Dur_99}, the main attention naturally
fell on its strongest form, the genuine multipartite entanglement, which is behind the multipartite tests of quantum nonlocality \cite{Greenberger_89}, complex behaviour of strongly correlated
systems \cite{Amico_08}, certain models of quantum computing \cite{Raussendorf_01} and increased precision of quantum measurements \cite{Giovannetti_04}.

So far, only examples of qubit states carrying genuine multipartite entanglement which can be verified solely from separable two-qubit reduced states (marginals) were
found \cite{Chen_14,Miklin_16} and demonstrated experimentally \cite{Micuda_19}. In all these cases, the set of marginals used to certify the entanglement comprised
{\it all} two-qubit marginals. Interestingly, genuine multipartite entanglement can be detected even from a smaller set of separable marginals. Indeed, multiqubit
states can be found possessing all two-qubit marginals separable, whose genuine multipartite entanglement can be inferred only from the so called
minimal set of two-qubit marginals \cite{Paraschiv_17}. The minimal set covers any part of the entire system and it contains only marginals between
nearest-neighbours, which guarantees that knowledge of the set suffices to confirm global entanglement. In geometric terms, if we
represent parts of the global state as vertices of a graph \cite{Steinbach_99} and the bipartite marginals as its edges, then the minimal set
corresponds to a tree-like graph. States with genuine multipartite entanglement which can be confirmed using only the elements of the minimal set
where found for all configurations of up to six qubits using the iterative numerical search algorithm \cite{Paraschiv_17} combining the machinery of entanglement
witnesses \cite{Horodecki_96,Terhal_00} with the tools of semi-definite programming \cite{Vandenbergehe_96}. The best example obtained 
was three-qubit \cite{Miklin_16} with the lowest witness mean being roughly
three times smaller than the witness mean for the scenario
in which all two-qubit marginals are known \cite{Miklin_16,Micuda_19}. Moreover, the difference is even more pronounced compared to
other theoretically predicted witness means \cite{Acin_01} of already successfully implemented multipartite entanglement witnesses experiments
\cite{Bourennane_04}. This indicates the complexity of the possible experimental demonstration of the studied effect.

\begin{figure}[ht]
\begin{center}
\includegraphics[width=0.40\textwidth]{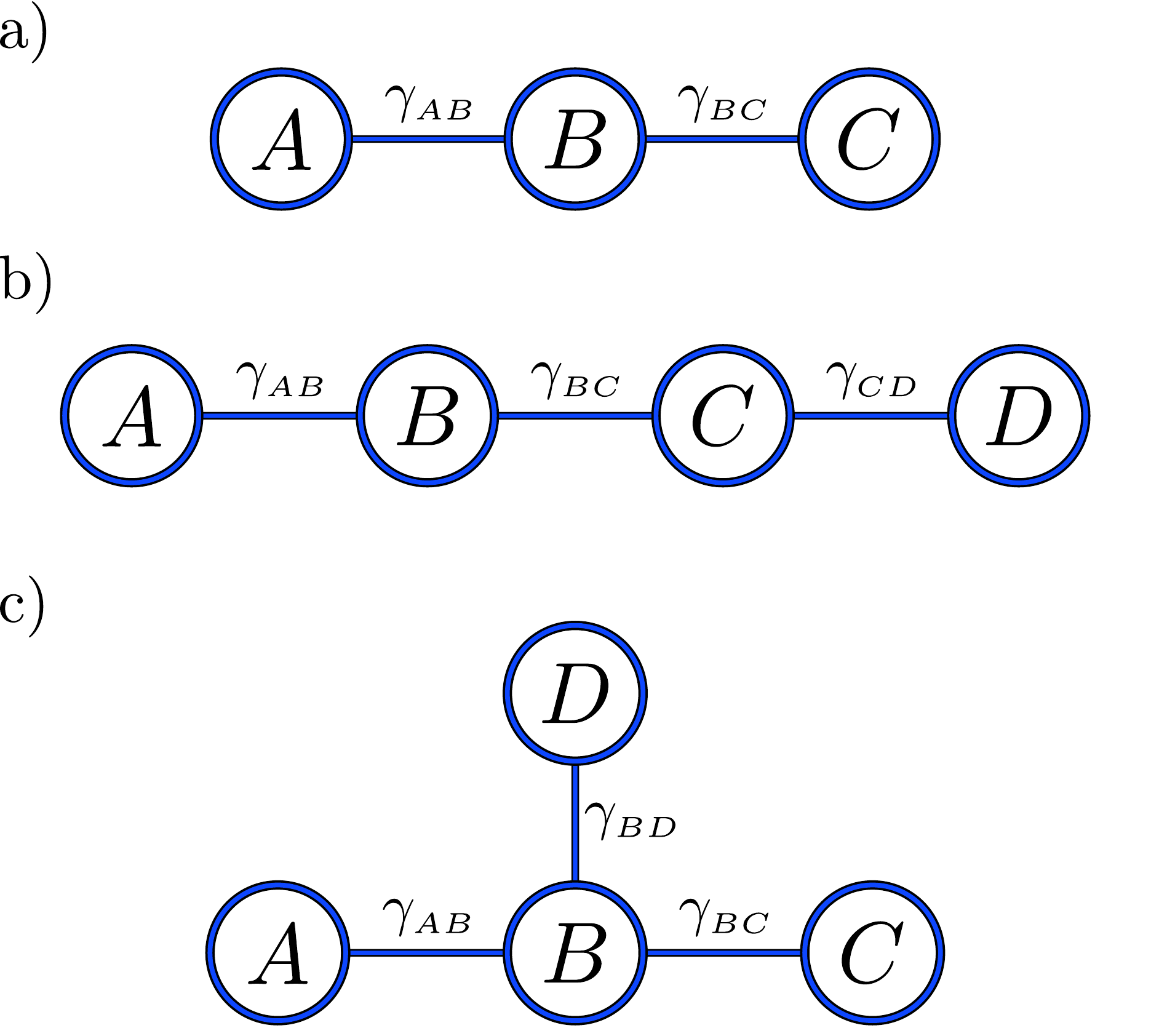}
\caption{Graphical representation of minimal sets of marginal CMs used to detect genuine multipartite entanglement for three and four modes. See text for details.}
\label{fig1}
\end{center}
\end{figure}

In this paper we take a different approach to the problem by seeking states with the investigated property
in the realm of Gaussian states \cite{Braunstein_05a,Weedbrook_12}. More precisely, we look for Gaussian
states with all two-mode marginals separable and whose genuine multipartite entanglement can be proved only from the minimal set of the
marginals. For this purpose, we use the methods of Gaussian multipartite entanglement witnesses \cite{Hyllus_06} to assemble a
Gaussian analog of the search algorithm \cite{Paraschiv_17}. By running the
algorithm we then find examples of the studied
states for all configurations of up to four modes (see Fig.~\ref{fig1}). Our
simplest examples involve only three modes similarly to
the simplest known qubit example which consists of three qubits \cite{Miklin_16}. The three-mode example gives, for the Gaussian
analog of the genuine multipartite entanglement witness mean, a value which is roughly of the same size as the theoretically predicted
values \cite{Hyllus_06} for already realized similar Gaussian multimode entanglement experiments \cite{DiGuglielmo_11}. Further, the required squeezing is less than one third of a vacuum unit.
Given the promising role of Gaussian states in the current problem, we also propose a linear-optical circuit for preparation
of the three-mode state, which is based on interference of three correlatively displaced squeezed beams on three beam splitters.
Our results reveal that a minimal set of overlapping separable marginals may suffice to reveal genuine multipartite entanglement
also in Gaussian scenario. Besides, they indicate that Gaussian continuous variables represent a promising alternative
platform for experimental demonstration of the studied property of genuine multipartite entanglement.

%%%%%%%%%%%%%%%%%%%%%%%%%%%%%%%%%%%%%%%%%%%%%%%%%%%%%%%%%%%%%%%%%%%%%%%%%%%%%%%%%%%%%%%%%%
\section{GAUSSIAN STATES}\label{sec_Gaussian_states}

The scene of our considerations is the set of Gaussian states
of systems with infinite-dimensional Hilbert space, which we
shall call modes in what follows. A collection of $N$ modes $A_{j}$, $j=1,2,\ldots,N$,
can be characterized by a vector $\xi=(x_{A_1},p_{A_1},\ldots,x_{A_N},p_{A_N})^{T}$ of
position and momentum quadratures $x_{A_{j}}$ and $p_{A_{j}}$, respectively,
which obey the canonical commutation rules $[\xi_j,\xi_k]=i(\Omega_{N})_{jk}$ with
$\Omega_{N}=\oplus_{j=1}^{N}i\sigma_{y}$, where
$\sigma_{y}$ is the Pauli-$y$ matrix.
%%%%%%%%%%%%%%%%%%%%%%%%%%%%%%%%%%%%%%%%%%%%%%%%%%%%%%%%%%%%%%%%%%%%%%%%%%%%%%%%%%%%
%\begin{equation}\label{OmegaN}
%\Omega_{N}=\bigoplus_{i=1}^{N}\left(\begin{array}{cc}
%0 & 1 \\
%-1 & 0\\
%\end{array}\right),
%\end{equation}
%%%%%%%%%%%%%%%%%%%%%%%%%%%%%%%%%%%%%%%%%%%%%%%%%%%%%%%%%%%%%%%%%%%%%%%%%%%%%%%%%%%%%%%%%%%%%%%%%%%%%
Gaussian states are defined as states with a Gaussian-shaped
phase-space Wigner function. An $N$-mode Gaussian state $\rho$ is thus fully described by a $2N\times 1$ vector
$\langle\xi\rangle=\mbox{Tr}[\xi\rho]$ of first moments and by a $2N\times 2N$ covariance matrix
(CM) $\gamma$ with entries $(\gamma)_{jk}=\langle\xi_{j}\xi_k+\xi_k\xi_j\rangle-2\langle\xi_{j}\rangle\langle\xi_k\rangle$.
The first moments can be nullified by local displacements and thus they are irrelevant as far as
the correlation properties investigated here are concerned. For this reason we set them to zero from now on.

Any CM $\gamma$ reflects the uncertainty principle by satisfying the inequality
%%%%%%%%%%%%%%%%%%%%%%%%%%%%%%%%%%%%%%%%%%%%%%%%%%%%%%%%%%%%%%%%%%%%%%%%%%%%%%%%%%%%%%%%%%%%%%%%%%%%%%%%%%
\begin{eqnarray}\label{Heisenbergg}
\gamma+i\Omega_{N}\geq0,
\end{eqnarray}
%%%%%%%%%%%%%%%%%%%%%%%%%%%%%%%%%%%%%%%%%%%%%%%%%%%%%%%%%%%%%%%%%%%%%%%%%%%%%%%%%%%%%%%%%%%%%%%%%%%%%%%%%%%
which is not only a necessary but also a sufficient condition for a real symmetric $2N\times 2N$
matrix $\gamma$ to be a CM of a physical quantum state \cite{Simon_00}. Besides, a CM also carries complete
information about the separability properties of the corresponding Gaussian state.
Recall first, that a quantum state $\rho_{jk}$ of two subsystems $j$ and $k$ is separable
if it can be expressed as a convex mixture of product states $\rho_{j|k}^{\rm sep}\equiv\sum_{i}p_{i}\rho_{j}^{(i)}\otimes\rho_{k}^{(i)}$,
where $\rho_{j}^{(i)}$ and $\rho_{k}^{(i)}$ are local states of subsystems $j$ and $k$, respectively.
If the state cannot be written in this form it is called entangled. Separability of a two-mode Gaussian state $\rho_{jk}$
can be ascertained by the positive partial transposition (PPT) criterion \cite{Peres_96,Horodecki_96,Simon_00}. On the CM level
the partial transposition operation $T_{j}$ with respect to mode $j$ transforms the CM $\gamma_{jk}$ of
the state as $\gamma_{jk}^{(T_{j})}=(\sigma_{z}\oplus\openone)\gamma_{jk}(\sigma_{z}\oplus\openone)$,
where $\sigma_{z}$ is the Pauli-$z$ matrix and $\openone$ is the $2\times 2$ identity matrix. The PPT criterion then says \cite{Simon_00},
that the state $\rho_{jk}$ is separable if and only if (iff) the matrix $\gamma_{jk}^{(T_{j})}$ is a physical CM, i.e., iff
%%%%%%%%%%%%%%%%%%%%%%%%%%%%%%%%%%%%%%%%%%%%%%%%%%%%%%%%%%%%%%%%%%%%%%%%%%%%%%%%%%%%%%%%%%%%%%%%%%%%%%%%%%%%%%%%%%%%%%%
\begin{equation}\label{HeisenbergTj}
\gamma_{jk}^{(T_{j})}+i\Omega_{2}\geq0.
\end{equation}
%%%%%%%%%%%%%%%%%%%%%%%%%%%%%%%%%%%%%%%%%%%%%%%%%%%%%%%%%%%%%%%%%%%%%%%%%%%%%%%%%%%%%%%%%%%%%%%%%%%%%%%%%%%%%%%%%%%%%%%

The PPT criterion is a sufficient condition for separability only for two-mode \cite{Simon_00} and
$1\times M$-mode \cite{Werner_01} Gaussian states. For systems where each party holds more than one
mode, one has to use a more powerful criterion \cite{Werner_01} according to which an $N$-mode Gaussian state
with CM $\gamma$ consisting of an $l$-mode subsystem $A \equiv A_{1}A_{2}\ldots A_{l}$
and an $(N-l)$-mode subsystem $B \equiv A_{l+1}A_{l+2}\ldots A_{N}$, is separable iff there are
CMs $\gamma_{A}$ and $\gamma_{B}$ of the subsystems such, that
%%%%%%%%%%%%%%%%%%%%%%%%%%%%%%%%%%%%%%%%%%%%%%%%%%%%%%%%%%%%%%%%%%%%%%%%%%%%%%%%%%%%%%%%%%
\begin{equation}\label{CMseparability}
\gamma-\gamma_{A}\oplus\gamma_{B}\geq0.
\end{equation}
%%%%%%%%%%%%%%%%%%%%%%%%%%%%%%%%%%%%%%%%%%%%%%%%%%%%%%%%%%%%%%%%%%%%%%%%%%%%%%%%%%%%%%%%%
The separability criterion (\ref{CMseparability}) is advantageous because it can be formulated as the following
semi-definite programme (SDP) \cite{Hyllus_06}:
%%%%%%%%%%%%%%%%%%%%%%%%%%%%%%%%%%%%%%%%%%%%%%%%%%%%%%%%%%%%%%%%%%%%%%%%%%%%%%%%%%%%%%%%%
\begin{equation}\label{sepprimal}
\begin{aligned}
& \underset{\gamma_{A},\gamma_{B},x_{e}}{\text{minimize}}
& & (-x_{e}) \\
& \text{subject to}
& &\gamma-\gamma_{A}\oplus\gamma_{B}\geq0,\\
& & &\gamma_{A}\oplus\gamma_{B}+(1+x_{e})i\Omega_{N}\geq0.
\end{aligned}
\end{equation}
%%%%%%%%%%%%%%%%%%%%%%%%%%%%%%%%%%%%%%%%%%%%%%%%%%%%%%%%%%%%%%%%%%%%%%%%%%%%%%%%%%%%%%%%%
If there is an optimal solution $x_{e}\geq0$, then CM $\gamma$ describes a separable state because there exist CMs
$\gamma_{A}$ and $\gamma_{B}$ such that the separability criterion (\ref{CMseparability}) is satisfied. If, on the other
hand, $x_{e}<0$, then the state with CM $\gamma$ is entangled.

\section{GAUSSIAN ENTANGLEMENT WITNESSES}\label{sec_Gaussian_entanglement_witnesses}

In practice, one needs most often to certify the presence of entanglement in a given state
rather than to show that it is separable. However, many entanglement criteria, including the PPT criterion
or criterion (\ref{CMseparability}) require knowledge of the entire quantum state and thus they are not
economical as far as the number of measurements is concerned. This also implies that the criteria
cannot be used in cases when we have access only to a part of the investigated state.
Nevertheless, it is still possible to detect entanglement provided that we have some a priori information
about the state. Namely, one can prove the presence of entanglement by measuring the so-called entanglement
witnesses \cite{Horodecki_96,Terhal_00}, which requires fewer measurements compared to the measurement of
the whole quantum state \cite{Guhne_02}.

\subsection{Bipartite entanglement witnesses}\label{subsec_Bipartite_witnesses}

For a bipartite state an entanglement witness is a Hermitian operator with a non-negative average for all separable states and a negative average on at least
one entangled state. However, the task of finding an entanglement witnesses for density matrices of continuous-variable modes is often hardly tractable
owing to their infinite dimension. A much more simple option, which is particularly suitable for Gaussian states,
is to seek entanglement witnesses for CMs \cite{Hyllus_06}. Such a witness is, for an $N$-mode state, represented by
a $2N\times2N$ real, symmetric and positive-semidefinite matrix $Z$, which satisfies the following conditions:
%%%%%%%%%%%%%%%%%%%%%%%%%%%%%%%%%%%%%%%%%%%%%%%%%%%%%%%%%%%%%%%%%%%%%%%%%%%%%%%%%%%%%%%%%%%%%%%%%%%%%%%%%%%%%%%%%%%%%
\begin{eqnarray}\label{Zconditions}
& \mbox{(i)} & \mbox{Tr}[\gamma Z]\geq 1,\quad \mbox{for all separable}\quad \gamma,\nonumber\\
& \mbox{(ii)} & \mbox{Tr}[\gamma Z]< 1,\quad \mbox{for some entangled}\quad \gamma.
\end{eqnarray}
%%%%%%%%%%%%%%%%%%%%%%%%%%%%%%%%%%%%%%%%%%%%%%%%%%%%%%%%%%%%%%%%%%%%%%%%%%%%%%%%%%%%%%%%%%%%%%%%%%%%%%%%%%%%%%%%%%%%%

Entanglement detection by means of matrix $Z$ possesses several advantages. First, the expression $\mbox{Tr}[\gamma Z]$ is a linear function of second moments and therefore it
can be measured by local homodyne detections followed by a suitable processing of the output photocurrents. More importantly, the expression also typically contains
only some elements of CM $\gamma$ and thus it requires fewer measurements than one needs to measure the entire CM. Another advantage of using the matrix
$Z$ is that for a given CM $\gamma$ it can be found numerically by solving the dual to program (\ref{sepprimal}) \cite{Hyllus_06}:
%%%%%%%%%%%%%%%%%%%%%%%%%%%%%%%%%%%%%%%%%%%%%%%%%%%%%%%%%%%%%%%%%%%%%%%%%%%%%%%%%%%%%%%%%
\begin{equation}\label{sepdual}
\begin{aligned}
& \underset{X_{1},X_{2}}{\text{minimize}}
& & \mbox{Tr}[\gamma X_{1}^{\rm re}]-1, \\
& \text{subject to}
& & X_{1}^{\rm bd, re}=X_{2}^{\rm bd, re},\quad X_{1}\geq0,\quad X_{2}\geq 0,\\
& & & \mbox{Tr}[i\Omega_{N} X_{2}]=-1.
\end{aligned}
\end{equation}
%%%%%%%%%%%%%%%%%%%%%%%%%%%%%%%%%%%%%%%%%%%%%%%%%%%%%%%%%%%%%%%%%%%%%%%%%%%%%%%%%%%%%%%%%
Here $X_{j}$, $j=1,2$, are $2N\times 2N$ Hermitian matrices, the symbol $X_{j}^{\rm re}$
stands for the real part of the matrix $X_{j}$, and $X_{j}^{\rm bd}=X_{j A}\oplus X_{j B}$, where $X_{j A}$ and $X_{j B}$
are diagonal blocks of the matrix $X_{j}$ corresponding to subsystems $A$ and $B$, respectively.

It can be shown \cite{Hyllus_06}, that for every feasible solution $X_{1}\oplus X_{2}$, the
matrix $X_{1}^{\rm re}$ satisfies
%%%%%%%%%%%%%%%%%%%%%%%%%%%%%%%%%%%%%%%%%%%%%%%%%%%%%%%%%%%%%%%%%%%%%%%%%%%%%%%%%%%%%%%%%%%%%%%%%%%%%
\begin{equation}\label{Wcondition1}
\mbox{Tr}[\gamma X_{1}^{\rm re}]\geq 1
\end{equation}
%%%%%%%%%%%%%%%%%%%%%%%%%%%%%%%%%%%%%%%%%%%%%%%%%%%%%%%%%%%%%%%%%%%%%%%%%%%%%%%%%%%%%%%%%%%%%%%%%%%%%
for every CM $\gamma$ of a separable state. Further, if $\gamma$ is a CM of an entangled state, then
%%%%%%%%%%%%%%%%%%%%%%%%%%%%%%%%%%%%%%%%%%%%%%%%%%%%%%%%%%%%%%%%%%%%%%%%%%%%%%%%%%%%%%%%%%%%%%%%%%%%%
\begin{equation}\label{Wcondition2}
\mbox{Tr}[\gamma X_{1}^{\rm re}]<1.
\end{equation}
%%%%%%%%%%%%%%%%%%%%%%%%%%%%%%%%%%%%%%%%%%%%%%%%%%%%%%%%%%%%%%%%%%%%%%%%%%%%%%%%%%%%%%%%%%%%%%%%%%%%%
This implies, that the real matrix $X_{1}^{\rm re}$ is an entanglement witness which is, in addition,
optimal in the sense that it yields the minimal value of $\mbox{Tr}[\gamma Z]$ out of all possible witnesses $Z$.
Needless to say, by adding more constraints into the SDP (\ref{sepdual}), one can seek witnesses with a special structure.
Below we will see, for instance, that one can seek witnesses which are `blind' to certain parts of CM $\gamma$.

\subsection{Genuine multipartite entanglement witnesses}\label{subsec_GME_witnesses}

Bipartite entanglement is just one particular kind of entanglement. In multipartite systems consisting of $N>2$ subsystems one
can investigate also multipartite entanglement, which occurs among more than two groups of subsystems. In general, it is possible
to split all subsystems into $k$ disjoint subsets and analyze entanglement with respect to the $k$-partite split.
We say that a state is $k$-separable if it is fully separable with respect to the $k$-partite split, i.e., if it
can be expressed as a convex mixture of product states with respect to the split. Otherwise, it is called entangled with
respect to the split. This allows us to classify multipartite states according to their separability properties with respect
to all possible $k$-partite splits for all possible $k$ \cite{Dur_99,Dur_00}. At the top of the hierarchy, there are
fully inseparable states which are not separable with respect to any $k$-partite split. Nevertheless,
even fully inseparable states in general do not carry the strongest form of multipartite entanglement. Namely,
some of them can be created by convex mixing of some $k$-separable states \cite{Guhne_09} and thus their preparation does not
require a collective operation on all subsystems as we would expect from truly multipartite entangled states.
For this reason, the concept of genuine $N$-partite entangled states was introduced as states that cannot be expressed as a
convex mixture of some $k$-separable states for any $k\geq2$ \cite{Acin_01}. Note, that any $k$-separable state with
$k>2$ is also $2$-separable. Consequently, a set of states that can be expressed as a convex mixture of some $k$-separable
states is a subset of the set of states that can be expressed as a convex mixture of some $2$-separable states, which are
fittingly called biseparable states. This reveals that for the presence of genuine multipartite entanglement in
a given quantum state it is sufficient to show that it is not biseparable.

The concept of biseparability carries over straightforwardly to CMs of $N$-mode Gaussian states.
For this purpose, let us collect modes $A_{j}$, $j=1,2,\ldots,N$, into the set $\mathcal{N}=\{A_{1},A_{2},\ldots,A_{N}\}$ and let $\mathcal{I}=\{1,2,\ldots,N\}$ be its index set.
Next, consider a nonempty proper index subset $\mathcal{J}_{k}=\{i_{1},i_{2},\ldots,i_{l}\}$ of $0<l<N$ elements of the index set $\mathcal{I}$ and let
$\bar{\mathcal{J}}_{k}=\mathcal{I}\backslash\mathcal{J}_{k}$ denotes its complement containing the remaining $N-l$ elements of $\mathcal{I}$. This allows
us to split the set $\mathcal{N}$ into $K\equiv 2^{N-1}-1$ different inequivalent $2$-partitions, called as bipartitions in what follows,
$\pi(k)\equiv\mathcal{M}_{\mathcal{J}_{k}}|\bar{\mathcal{M}}_{\mathcal{J}_{k}}$, $k=1,2,\ldots,K$, where
$\mathcal{M}_{\mathcal{J}_{k}}=\{A_{i_{1}},A_{i_{2}},\ldots,A_{i_{l}}\}$ and $\bar{\mathcal{M}}_{\mathcal{J}_{k}}=\mathcal{M}_{\bar{\mathcal{J}}_{k}}=\mathcal{N}\backslash\mathcal{M}_{\mathcal{J}_{k}}$.

Moving to the criterion of biseparability one can show \cite{Hyllus_06}, that an $N$-mode Gaussian state with CM $\gamma$ is
biseparable iff there exist bipartitions $\pi(k)$ and CMs $\gamma_{\pi(k)}$ which are block diagonal with respect to the bipartition $\pi(k)$, and probabilities $\lambda_{k}$ such that
%%%%%%%%%%%%%%%%%%%%%%%%%%%%%%%%%%%%%%%%%%%%%%%%%%%%%%%%%%%%%%%%%%%%%%%%%%%%%%%%%%%%%%%%%%%%%%%%%%%%%
\begin{equation}\label{biseparable}
\gamma-\sum_{k=1}^{K}\lambda_{k}\gamma_{\pi(k)}\geq0.
\end{equation}
%%%%%%%%%%%%%%%%%%%%%%%%%%%%%%%%%%%%%%%%%%%%%%%%%%%%%%%%%%%%%%%%%%%%%%%%%%%%%%%%%%%%%%%%%%%%%%%%%%%%%

Similarly as bipartite separability can be decided by solving the SDP (\ref{sepprimal}),
biseparability embodied by condition (\ref{biseparable}) can also be decided by solving an SDP \cite{Hyllus_06}.
Analogously, just like an optimal witness of bipartite entanglement can be obtained by solving the dual problem
(\ref{sepdual}) of the former SDP, the optimal witness of genuine $N$-partite entanglement can be found by solving the
dual problem of the corresponding SDP \cite{Hyllus_06}. Recall first, that the witness of genuine $N$-partite
entanglement is represented by a $2N\times 2N$ real, symmetric, and positive-semidefinite matrix $Z$ satisfying conditions \cite{Hyllus_06}
%%%%%%%%%%%%%%%%%%%%%%%%%%%%%%%%%%%%%%%%%%%%%%%%%%%%%%%%%%%%%%%%%%%%%%%%%%%%%%%%%%%%%%%%%%%%%%%%%%%%%%%%%%%%%%%%%%%%%
\begin{eqnarray}\label{ZGME}
& \mbox{(i)} & \mbox{Tr}[\gamma Z]\geq 1,\quad \mbox{for all biseparable}\quad \gamma,\nonumber\\
& \mbox{(ii)} & \mbox{Tr}[\gamma Z]< 1,\quad \mbox{for some entangled}\quad \gamma.
\end{eqnarray}
%%%%%%%%%%%%%%%%%%%%%%%%%%%%%%%%%%%%%%%%%%%%%%%%%%%%%%%%%%%%%%%%%%%%%%%%%%%%%%%%%%%%%%%%%%%%%%%%%%%%%%%%%%%%%%%%%%%%%
For a given CM $\gamma$ the witness can be found by solving the following dual problem \cite{Hyllus_06}:
%%%%%%%%%%%%%%%%%%%%%%%%%%%%%%%%%%%%%%%%%%%%%%%%%%%%%%%%%%%%%%%%%%%%%%%%%%%%%%%%%%%%%%%%%
\begin{widetext}
\begin{equation}\label{bisepdual}
\begin{aligned}
& \underset{X}{\text{minimize}}
& & \mbox{Tr}[\gamma X_{1}^{\mathrm{re}}]-1\\
& \text{subject to}
& &X_{1}^{{\rm re, bd}, \pi(k)}=X_{k+1}^{{\rm re, bd}, \pi(k)}\quad \mbox{for all}\quad k=1,\ldots,K,\\
& & &\mbox{Tr}[i\Omega_{N} X_{k+1}]+X_{K+2}-X_{K+3}+X_{K+3+k}=0,\quad \mbox{for all}\quad k=1,\ldots,K,\\
& & &X_{K+2}-X_{K+3}=1.
\end{aligned}
\end{equation}
\end{widetext}
%%%%%%%%%%%%%%%%%%%%%%%%%%%%%%%%%%%%%%%%%%%%%%%%%%%%%%%%%%%%%%%%%%%%%%%%%%%%%%%%%%%%%%%%%
Here, the minimization is preformed over Hermitian positive-semidefinite $[2N(K+1)+2+K]$-dimensional block-diagonal matrix
%%%%%%%%%%%%%%%%%%%%%%%%%%%%%%%%%%%%%%%%%%%%%%%%%%%%%%%%%%%%%%%%%%%%%%%%%%%%%%%%%%%%%%%%%
\begin{equation}\label{X}
X=\bigoplus_{j=1}^{2K+3}X_{j},
\end{equation}
%%%%%%%%%%%%%%%%%%%%%%%%%%%%%%%%%%%%%%%%%%%%%%%%%%%%%%%%%%%%%%%%%%%%%%%%%%%%%%%%%%%%%%%%%
with $X_{j}$, $j=1,2,\ldots,K+1$ being $2N\times 2N$ Hermitian matrices and $X_{j}$, $j=K+2,K+3,\ldots,2K+3$
being $1\times 1$ Hermitian matrices, i.e., real numbers. Further, the $k$-th equation $X_{1}^{{\rm re, bd}, \pi(k)}=X_{k+1}^{{\rm re, bd}, \pi(k)}$
imposes a constraint on diagonal blocks of the matrices $X_{1}$ and $X_{k+1}$ written in the block form with respect to the bipartition $\pi(k)$.
More precisely, let us express the matrix $X_{j}$ in the block-form with respect to the $N$-partite split $A_{1}|A_{2}|\ldots|A_{N}$,
%%%%%%%%%%%%%%%%%%%%%%%%%%%%%%%%%%%%%%%%%%%%%%%%%%%%%%%%%%%%%%%%%%%%%
\begin{eqnarray}\label{Xjblock}
X_{j}=\left(\begin{array}{cccc}
(X_{j})_{11} & (X_{j})_{12} & \ldots & (X_{j})_{1N} \\
(X_{j})_{12}^{\dag} & (X_{j})_{22} & \ldots & (X_{j})_{2N} \\
\vdots & \vdots & \ddots & \vdots\\
(X_{j})_{1N}^{\dag} & (X_{j})_{2N}^{\dag} & \ldots & (X_{j})_{NN}\\
\end{array}\right),
\end{eqnarray}
%%%%%%%%%%%%%%%%%%%%%%%%%%%%%%%%%%%%%%%%%%%%%%%%%%%%%%%%%%%%%%%%%%%%%%
where $(X_{j})_{mn}$ is a $2\times 2$ block. Then, the matrix $X_{j}^{\mathrm{bd}, \pi(k)}$ is of the same block form with the $2\times 2$
blocks given by
%%%%%%%%%%%%%%%%%%%%%%%%%%%%%%%%%%%%%%%%%%%%%%%%%%%%%%%%%%%%%%%%%%%%%%%%%%%%%%%%%%%%%%%%%%%%%%%%%%%%%%%%%%%%%%%%%%%%%%%%%%%%%%%%%%%%%%%%%%%%%%
\begin{equation}
\label{Xbd}
\left(X_{j}^{\mathrm{bd}, \pi(k)}\right)_{mn}=\left\{\begin{array}{lllll} (X_{j})_{mn}, & \!\textrm{if} & \hspace{-1.4cm} m,n\in\mathcal{J}_{k} & \textrm{or} & \bar{\mathcal{J}}_{k};\\
\mathbb{O}, & \textrm{otherwise}, & & &\\
\end{array}\right.
\end{equation}
%%%%%%%%%%%%%%%%%%%%%%%%%%%%%%%%%%%%%%%%%%%%%%%%%%%%%%%%%%%%%%%%%%%%%%%%%%%%%%%%%%%%%%%%%%%%%%%%%%%%%%%%%%%%%%%%%%%%%%%%%%%%%%%%%%%%%%%%%%%%%
where $\mathbb{O}$ is the $2\times 2$ zero matrix. For relevant cases $N=3$ and $N=4$ discussed in this paper an explicit form of the matrices
$X_{j}^{\mathrm{bd}, \pi(k)}$ can be found in Appendix~\ref{sec_app_I}.

According to the results of Ref.~\cite{Hyllus_06}, for every feasible solution $X$ of the dual program (\ref{bisepdual})
the matrix $X_{1}^{\mathrm{re}}$ is an optimal genuine multipartite entanglement witness.

\subsection{Blind genuine multipartite entanglement witnesses}\label{subsec_blind_GME_witnesses}

The witness obtained by solving the programme (\ref{bisepdual}) acts on the entire CM $\gamma$ and therefore enables us to certify genuine multipartite entanglement
provided that all elements of the CM are known. Viewed from a different perspective, it is equivalent to witnessing the entanglement from all two-mode marginal
CMs, because they completely determine the global CM. In this respect, the domain of Gaussian states differs from the qubit case, where the knowledge of all two-qubit marginals
is not generally equivalent to the knowledge of the whole density matrix. To make the task on inference of genuine multipartite entanglement from marginals
in Gaussian scenario meaningful, we thus have to work only with a proper subset of the set of all two-mode marginal CMs. In what follows, we utilize
the so-called minimal sets of bipartite marginals, which were introduced recently in Ref.~\cite{Paraschiv_17} to solve the task for qubits. Obviously, a necessary condition for the
set to allow detection of global entanglement is that it contains all modes and that one cannot divide it into a subset and its complement without having a common mode.
Among all such sets a particularly important role play further irreducible sets containing a minimum possible number of two-mode marginals.

A more convenient pictorial representation of such minimal sets was put forward in Ref.~\cite{Paraschiv_17} in the form on an unlabeled tree \cite{Steinbach_99}, which is a special form of an undirected
connected graph containing no cycles. Recall, that a graph is a pair $G=(V,E)$ of a set $V=\{1,2,\ldots,N\}$ of vertices and a set $E\subseteq K\equiv\{\{u,v\}|(u,v)\in V^2 \wedge u\ne v\}$
of edges \cite{West_01}. In our case a vertex $j$ of the graph represents mode $A_{j}$, whereas the edge connecting adjacent vertices $j$ and $k$ represents marginal CM $\gamma_{A_{j}A_{k}}$.
By definition, the minimal set contains two-mode marginal CMs corresponding to the edges in the respective tree denoted as $T=(V,E')$. A closed formula for the number
of non-isomorphic trees with $N$ vertices is not known, yet for small $N$ it can be found in Ref.~\cite{A000055}. In particular, all trees for the three-mode case ($N=3$) and
the four-mode case ($N=4$) are depicted in Fig.~\ref{fig1}, where we performed the following identification $A\equiv A_{1}, B\equiv A_{2}, C\equiv A_{3}$, and $D\equiv A_{4}$.

Ignorance of some sectors of CM $\gamma$ requires to impose some additional constraints onto the structure of the witness $X_{1}^{\mathrm{re}}$ being the solution of SDP (\ref{bisepdual}).
Specifically, as the respective tree is connected, the minimal set contains all single-mode CMs as well as $2\times2$ blocks of correlations between the
modes corresponding to the endpoints of the edges of the tree $T$. The part of the CM $\gamma$ which we do not know is therefore given by all $2\times2$ off-diagonal blocks of
correlations between pairs of modes carried by the marginal two-mode CMs contained in the complement of the minimal set. The elements of the complement correspond to the
edges in the complement graph $\bar{T}=(V,K\backslash E')$, i.e., to the edges which have to be added to the original tree $T$ to form the complete graph. Since for a given $N$ the complete graph
contains $\binom{N}{2}$ edges and the tree $T$ contains exactly $N-1$ edges \cite{West_01}, the number of unknown blocks of correlations is equal to $L\equiv(N-1)(N-2)/2$.
Further, as $\mbox{Tr}[\gamma X_{1}^{\rm re}]=\sum_{j,k}(\gamma)_{jk}(X_{1}^{\rm re})_{jk}$, in order for the witness
$X_{1}^{\mathrm{re}}$ not to act on the unknown blocks of CM $\gamma$, its blocks in places of the unknown blocks have to vanish. More precisely, if we
express the witness $X_{1}^{\mathrm{re}}$ in the block form with respect to $N$-partite split $A_{1}|A_{2}|\ldots|A_{N}$ similar to Eq.~(\ref{Xjblock}),
its $2\times 2$ off-diagonal blocks have to satisfy the following set of $L$ equations:
%%%%%%%%%%%%%%%%%%%%%%%%%%%%%%%%%%%%%%%%%%%%%%%%%%%%%%%%%%%%%%%%%%%%%%%%%%%%%%%%%%%%%%%%%%%%%%%%%%%
\begin{equation}\label{blind}
(X_{1}^{\rm re})_{mn}=\mathbb{O},\quad \mbox{if} \quad \{m,n\}\in K\backslash E',
\end{equation}
%%%%%%%%%%%%%%%%%%%%%%%%%%%%%%%%%%%%%%%%%%%%%%%%%%%%%%%%%%%%%%%%%%%%%%%%%%%%%%%%%%%%%%%%%%%%%%%%%%%
which have to be added to the SDP (\ref{bisepdual}) as additional constraints. For $N=3$ and the tree in Fig.~\ref{fig1}~a), the constraint reads explicitly as
%%%%%%%%%%%%%%%%%%%%%%%%%%%%%%%%%%%%%%%%%%%%%%%%%%%%%%%%%%%%%%%%%%%%%%%%%%%%%%%%%%%%%%%%%%%%%%%%%%%%%
\begin{eqnarray}\label{blind3}
(X_{1}^{\mathrm{re}})_{13}=\mathbb{O}.
\end{eqnarray}
%%%%%%%%%%%%%%%%%%%%%%%%%%%%%%%%%%%%%%%%%%%%%%%%%%%%%%%%%%%%%%%%%%%%%%%%%%%%%%%%%%%%%%%%%%%%%%%%%%%%%
Likewise, in the case $N=4$ and for the linear tree in Fig.~\ref{fig1}~b), the constraints are
%%%%%%%%%%%%%%%%%%%%%%%%%%%%%%%%%%%%%%%%%%%%%%%%%%%%%%%%%%%%%%%%%%%%%%%%%%%%%%%%%%%%%%%%%%%%%%%%%%%%%
\begin{eqnarray}\label{blind41}
(X_{1}^{\mathrm{re}})_{13}=(X_{1}^{\mathrm{re}})_{14}=(X_{1}^{\mathrm{re}})_{24}=\mathbb{O},
\end{eqnarray}
%%%%%%%%%%%%%%%%%%%%%%%%%%%%%%%%%%%%%%%%%%%%%%%%%%%%%%%%%%%%%%%%%%%%%%%%%%%%%%%%%%%%%%%%%%%%%%%%%%%%%
whereas for the `t'-shaped tree  in Fig.~\ref{fig1}~c) one gets the constraints of the following form:
%%%%%%%%%%%%%%%%%%%%%%%%%%%%%%%%%%%%%%%%%%%%%%%%%%%%%%%%%%%%%%%%%%%%%%%%%%%%%%%%%%%%%%%%%%%%%%%%%%%%%
\begin{eqnarray}\label{blind42}
(X_{1}^{\mathrm{re}})_{13}=(X_{1}^{\mathrm{re}})_{14}=(X_{1}^{\mathrm{re}})_{34}=\mathbb{O}.
\end{eqnarray}
%%%%%%%%%%%%%%%%%%%%%%%%%%%%%%%%%%%%%%%%%%%%%%%%%%%%%%%%%%%%%%%%%%%%%%%%%%%%%%%%%%%%%%%%%%%%%%%%%%%%%

\section{SEARCH ALGORITHM}\label{sec_search_algorithm}

The goal of the present paper is to find an example of a Gaussian state with all two-mode marginals separable and whose genuine multipartite entanglement can be verified
solely from the minimal set of two-mode marginals. Recently, multiqubit examples of such states have been found \cite{Paraschiv_17} using a two-step
algorithm proposed in Ref.~\cite{Miklin_16}. Here, we employ the following Gaussian analog of the algorithm:

{\it Step 0:} Generate a random pure Gaussian state with CM $\gamma_{0}$ which has, for simplicity, no $x-p$ correlations.

{\it Step 1:} For CM $\gamma_{0}$, find a witness $X_{1}^{\mathrm{re}}$ by solving numerically the SDP (\ref{bisepdual}) supplemented with the constraints (\ref{blind}), which we
shall call as the SDP~1. Note, that the SDP~1 can be solved by modifying the freely available routine \cite{Lofberg} in Matlab by adding the constraints (\ref{blind}) into it.

{\it Step 2:} Find a CM $\gamma$ that gives the least value of $\mbox{Tr}[\gamma X_{1}^{\mathrm{re}}]$ for the witness $X_{1}^{\mathrm{re}}$ from the
first step under the constraint that the CM possesses all two-mode marginals separable. Again, the search can be accomplished by solving the following SDP:
%%%%%%%%%%%%%%%%%%%%%%%%%%%%%%%%%%%%%%%%%%%%%%%%%%%%%%%%%%%%%%%%%%%%%%%%%%%%%%%%%%%%%%%%%
\begin{equation}\label{step2}
\begin{aligned}
& \underset{\gamma}{\text{minimize}}
& & \mbox{Tr}[\gamma X_{1}^{\mathrm{re}}] \\
& \text{subject to}
& &\gamma+i\Omega_{N}\geq0,\\
& & & \gamma_{jk}^{(T_{j})}+i\Omega_{2}\geq0, \quad \mbox{for all}\quad j\ne k=1,\ldots,N,\\
& & & (\gamma)_{2j-1,2k}=(\gamma)_{2j,2k-1}=0, \quad j,k=1,\ldots,N,
\end{aligned}
\end{equation}
%%%%%%%%%%%%%%%%%%%%%%%%%%%%%%%%%%%%%%%%%%%%%%%%%%%%%%%%%%%%%%%%%%%%%%%%%%%%%%%%%%%%%%%%%
which is called as SDP~2 from now. Here, we carry out the minimization over all real symmetric $2N\times 2N$ matrices $\gamma$.
The first constraint guarantees that the matrix $\gamma$ is a CM of a physical quantum state, whereas the second constraint assures that
all its two-mode marginal CMs $\gamma_{jk}$ are separable. Finally, due to the third constraint we perform minimization
only over matrices $\gamma$ which do not contain any $x-p$ correlations.

By putting the obtained solution from Step~2 as an input to Step~1 we can iteratively seek the CM with
the desired properties. In the next section we give explicit examples of such CMs for all three-mode and four-mode minimal sets.

\section{Results}\label{sec_results}

\subsection{Three modes}
\label{sec_results_3modes}

First, we did a numerical search of a three-mode example of the investigated
effect. Running SDP~1 and SDP~2 successively for 10 iterations for $N=3$,
we found several examples of states with all two-mode marginals separable and
whose genuine three-mode entanglement can be verified solely from the nearest
neighbour marginal CMs $\gamma_{AB}$ and $\gamma_{BC}$ (see Fig.~\ref{fig1}~a)).
The CMs typically exhibited large diagonal entries and required high squeezing
for preparation. To get experimentally easier accessible CM, we therefore added another two
constraints to the SDP~2 (\ref{step2}). First, we limited the diagonal elements of
the CM to lie within the range $[1,10]$ and second, we also constrained the smallest
eigenvalue of the sought CM $\gamma$ to be above $0.2$. The best CM we got
in this way giving the least value of $\mbox{Tr}[\gamma Z]$ reads after the
rounding to two decimal places as

%%%%%%%%%%%%%%%%%%%%%%%%%%%%%%%%%%%%%%%%%%%%%%%%%%%%%%%%%%%%%%%%%%%%%
\begin{eqnarray*}\label{gamma3}
\gamma_{3} =\left(\begin{array}{cccccc}
    1.34   &       0 & -0.35   &       0 & -0.82   &       0\\
         0 & 10.00   &       0 &  8.45   &       0 &  1.87  \\
   -0.35   &       0 &  7.8  0 &       0 & -8.05   &       0\\
         0 &  8.45   &       0 &  7.92   &       0 &  2.09  \\
   -0.82   &       0 & -8.05   &       0 & 10.00   &       0\\
         0 &  1.87   &       0 &  2.09   &       0 &  1.62
\end{array}\right),\nonumber\\
\end{eqnarray*}
%%%%%%%%%%%%%%%%%%%%%%%%%%%%%%%%%%%%%%%%%%%%%%%%%%%%%%%%%%%%%%%%%%%%%%
and by running the SDP~1 for the rounded CM $\gamma_{3}$ we got
$\mbox{Tr}[\gamma_{3} Z_{3}]-1\doteq-0.143$. The corresponding witness, which is
blind to the correlations between a pair of modes $(A,C)$, is after rounding to three
decimal places given by
%%%%%%%%%%%%%%%%%%%%%%%%%%%%%%%%%%%%%%%%%%%%%%%%%%%%%%%%%%%%%%%%%%%%%%
\begin{eqnarray*}\label{Z}
Z_{3} \!\! = \!\! 10^{-2} \!\!
\left( \!\! \begin{array}{cccccc}
    6.8 &      0 &  -0.4 &      0 &      0 &     0\\
       0 &  34.3 &      0 & -39.5 &      0 &     0\\
   -0.4 &      0 &  25.1 &      0 &  20.9 &     0\\
       0 & -39.5 &      0 &  46.1 &      0 & -2.0\\
       0 &      0 &  20.9 &      0 &  17.5 &     0\\
       0 &      0 &      0 &  -2.0 &      0 &  6.6
\end{array} \!\! \right) \!\!.\nonumber\\
\end{eqnarray*}
%%%%%%%%%%%%%%%%%%%%%%%%%%%%%%%%%%%%%%%%%%%%%%%%%%%%%%%%%%%%%%%%%%%%%%
%%%%%%%%%%%%%%%%%%%%%%%%%%%%%%%%%%%%%%%%%%%%%%%%%%%%%%%%%%%%%%%%%%%%%%

The separability of all marginals is evidenced by Tab.~\ref{table1}
%%%%%%%%%%%%%%%%%%%%%%%%%%%%%%%%%%%%%%%%%%%%%%%%%%%%%%%%%%%%%%%%%%%%%%%%%%%%%%%%%%%%%%%%%%%%%%%%%%%%%%

\begin{table}[ht]
\caption{Minimal eigenvalue $\varepsilon_{jk}\equiv\mathrm{min}\{\mathrm{eig}[\gamma_{3,jk}^{(T_{j})}+i\Omega_{2}]\}$.} \centering
\begin{tabular}{| c | c | c | c |}
\hline $jk$                 & AB        & AC         & BC      \\
\hline $\varepsilon_{jk}$   & 0.002     & 0.849      & 0.004    \\
\hline
\end{tabular}
\label{table1}
\end{table}

Inspection of Tab.~\ref{table1} reveals that all eigenvalues are strictly
positive and therefore all three two-mode marginal states are separable by PPT
criterion as required.

The present result can be compared with the results for qubits derived in
Ref.~\cite{Paraschiv_17}. 
Note first that the value of $\mbox{Tr}[\gamma_{3} Z_{3}]-1\doteq-0.143$ found
here, for the simplest three-mode state, is slightly larger than the theoretical value of $-0.103$ for the same 
quantity of the comparable effect of Gaussian bound entanglement \cite{Hyllus_06,Werner_01}, 
which was already observed experimentally \cite{DiGuglielmo_11}. On the other hand, 
the best qubit mean of $\mbox{Tr}[\rho W]\doteq-6.58\cdot 10^{-3}$ obtained for the  
three-qubit state \cite{Miklin_16} is approximately three times smaller 
than the best theoretical witness mean of $-1.98\cdot 10^{-2}$ for the 
case when all two-qubit marginals are known \cite{Miklin_16}, which was recently 
demonstrated in \cite{Micuda_19}. Recall further, that in the qubit scenario 
the noise tolerance is $5\%$ \cite{Miklin_16}. For comparison, 
the produced state with CM $\gamma_{3}$ also tolerates the addition of a small amount of thermal noise, 
i.e., the CM $\gamma_{p}=\gamma_{3}+p\openone$ exhibits the effect for up to $p\doteq0.1$, 
yet the value is the same as one would get for the successfully demonstrated Gaussian bound 
entanglement \cite{Hyllus_06}. 

All these facts indicate the domain of Gaussian states to be a more promising platform for 
the near-future experimental demonstration of the analyzed effect. Therefore, in the next section we present a 
linear-optical scheme for preparation of a close approximation of the state with CM $\gamma_{3}$. 
However, before doing so, we first construct also four-mode states carrying the investigated property.

%%%%%%%%%%%%%%%%%%%%%%%%%%%%%%%%%%%%%%%%%%%%%%%%%%%%%%%%%%%%%%%%%%%%%%%%%%%%%%%%%%%%%%%%%%%%%%%%%%%
\subsection{Four modes}

Next, we extended the search of example CMs to four modes. In this case there are two different minimal sets
of marginals corresponding to the linear tree and the `t'-shaped tree displayed in Figs.~\ref{fig1}~b) and c),
respectively. Through the same procedure as for the three-mode case, we found CMs with the desired
properties for both the minimal sets, which are given explicitly below.

%%%%%%%%%%%%%%%%%%%%%%%%%%%%%%%%%%%%%%%%%%%%%%%%%%%%%%%%%%%%%%%%%%%%%%%%
\subsubsection{Linear tree}

First, we considered the minimal set of marginals given by the CMs
$\gamma_{AB}$, $\gamma_{BC}$ and $\gamma_{CD}$, corresponding to the edges in the linear tree in
Fig.~\ref{fig1}~b). This was reflected by inclusion of the constraints (\ref{blind41}) into our
search algorithm. By running the algorithm for 10 iterations, we produced several four-mode
CMs with the desired properties. The best such CM is

\begin{widetext}
\begin{eqnarray}\label{gammaLinear}
\gamma_{4}^{(1)}=\left(\begin{array}{cccccccc}
    2.83 &     0 & -0.02 &     0 & -1.38 &     0 &  2.83 &     0\\
       0 &  7.18 &     0 &  8.06 &     0 &  7.09 &     0 & -4.12\\
   -0.02 &     0 &  3.91 &     0 & -2.46 &     0 &  4.73 &     0\\
       0 &  8.06 &     0 &  9.79 &     0 &  8.47 &     0 & -4.81\\
   -1.38 &     0 & -2.46 &     0 &  2.58 &     0 & -4.68 &     0\\
       0 &  7.09 &     0 &  8.47 &     0 & 10.00 &     0 & -3.08\\
    2.83 &     0 &  4.73 &     0 & -4.68 &     0 & 10.00 &     0\\
       0 & -4.12 &     0 & -4.81 &     0 & -3.08 &     0 &  3.22
\end{array}\right).
\end{eqnarray}
\end{widetext}

The optimal witness $Z_{4}^{(1)}$, which is blind to correlations of modes $(A,C), (A,D)$ and $(B,D)$, gives the value of
$\mbox{Tr}[\gamma^{(1)}_4 Z_4^{(1)}]-1\doteq-0.069$ and it can be found in Appendix~\ref{sec_app_II}. The separability of all marginals can be confirmed
again by the PPT criterion (\ref{HeisenbergTj}) which is captured in Tab.~\ref{tableLinear}.

%%%%%%%%%%%%%%%%%%%%%%%%%%%%%%%%%%%%%%%%%%%%%%%%%%%%%%%%%%%%%%%%%%%%%%%%%%%%%%%%%%%%%%%%%%%%%%%%%%%%%%
\begin{table}[ht!]
\caption{Minimal eigenvalue $\varepsilon_{jk}^{(1)}\equiv\mbox{min}\{\mbox{eig}[(\gamma_{4,jk}^{(1)})^{(T_j)} +
    i \Omega_2]$.}
\centering
\begin{tabular}{| c | c | c | c | c | c | c |}
\hline $jk$ & AB & AC & AD & BC & BD & CD \\
\hline $\varepsilon_{jk}^{(1)}$ & 0.005 & 0.347 & 0.213 & 0.004 & 0.087 & 0.224 \\
\hline
\end{tabular}
\label{tableLinear}
\end{table}
%%%%%%%%%%%%%%%%%%%%%%%%%%%%%%%%%%%%%%%%%%%%%%%%%%%%%%%%%%%%%%%%%%

As all entries in the second row of the Tab.~\ref{tableLinear} are strictly positive, all two-mode marginal CMs of CM $\gamma_{4}^{(1)}$
are separable as required. Note further, that the effect is roughly half that of the three-mode case,
which makes its experimental demonstration a bigger challenge. 

%%%%%%%%%%%%%%%%%%%%%%%%%%%%%%%%%%%%%%%%%%%%%%%%%%%%%%%%%%%%%%%%%%%%%%%%%%
\subsubsection{`t'-shaped tree}

Finally, we looked for states whose genuine four-mode entanglement can be witnessed
from its nearest-neighbour marginals as per the graph in Fig.~\ref{fig1} c).
This corresponds to the `t'-shaped tree for which the minimal set comprise
marginal CMs  $\gamma_{AB}, \gamma_{BC}$ and $\gamma_{BD}$, and the
witness then fulfils the constraints (\ref{blind42}).
The best example CM found reads as
%%%%%%%%%%%%%%%%%%%%%%%%%%%%%%%%%%%%%%%%%%%%%%%%%%%%%%%%%%%%%%%%%%%%%%%%%%%%%%%%%%%%%%%%%
\begin{widetext}
\begin{eqnarray}\label{gammaTshape}
\gamma_{4}^{(2)}=\left(\begin{array}{cccccccc}
    5.23   &       0 &  0.45   &       0 & -0.02   &       0 & -2.43   &      0\\
         0 &  1.16   &       0 &  3.00   &       0 &  1.15   &       0 &  0.51 \\
    0.45   &       0 &  3.35   &       0 &  0.91   &       0 & -5.20   &      0\\
         0 &  3.00   &       0 & 10.00   &       0 &  3.52   &       0 &  2.06 \\
   -0.02   &       0 &  0.91   &       0 &  4.09   &       0 & -2.97   &      0\\
         0 &  1.15   &       0 &  3.52   &       0 &  1.62   &       0 &  0.62 \\
   -2.43   &       0 & -5.20   &       0 & -2.97   &       0 & 10.00   &      0\\
         0 &  0.51   &       0 &  2.06   &       0 &  0.62   &       0 &  1.49
\end{array}\right).
\end{eqnarray}
\end{widetext}

The corresponding optimal witness $Z^{(2)}_4$ is blind to intermodal correlations of pairs of modes $(A,C), (A,D)$ and $(C,D)$. It gives the value of
$\mbox{Tr}[\gamma^{(2)}_4 Z^{(2)}_4]-1\doteq-0.068$ and its explicit form can be found in Appendix~\ref{sec_app_II}. Once again, the separability of the
marginals can be verified via the PPT criterion. The results are summarized in Tab.~\ref{tableTshape}.

%%%%%%%%%%%%%%%%%%%%%%%%%%%%%%%%%%%%%%%%%%%%%%%%%%%%%%%%%%%%%%%%%%%%%%%%%%%%%%%%%%%%%%%%%%%%%%%%%%%%%%
\begin{table}[ht]
    \caption{Minimal eigenvalue $\varepsilon_{jk}^{(2)}\equiv\mbox{min}\{\mbox{eig}[(\gamma_{4,jk}^{(2)})^{(T_j)} + i \Omega_2]\}$.}
\centering
\begin{tabular}{| c | c | c | c | c | c | c |}
\hline $jk$ & AB & AC & AD & BC & BD & CD \\
\hline $\varepsilon_{jk}^{(2)}$
& 0.0481 & 0.0032 & 0.5256 & 0.1103 & 0.0001 & 0.5489 \\
\hline
\end{tabular}
\label{tableTshape}
\end{table}
%%%%%%%%%%%%%%%%%%%%%%%%%%%%%%%%%%%%%%%%%%%%%%%%%%%%%%%%%%%%%%%%%%

The effect is about the same strength as for the linear tree.
A point to note is that, for qubits, a pure state example was found for
the `t'-shaped tree in Ref.~\cite{Paraschiv_17} while we found only a
mixed-state example in the Gaussian scenario. 

%%%%%%%%%%%%%%%%%%%%%%%%%%%%%%%%%%%%%%%%%%%%%%%%%%%%%%%%%%%%%%%%%%%%%%%%%%%%%%%%%%%%%%%%%%%%%%%%%%%%%%

\section{Experimental scheme}\label{sec_scheme}

\begin{figure}[ht]
\begin{center}
\includegraphics[width=0.48\textwidth]{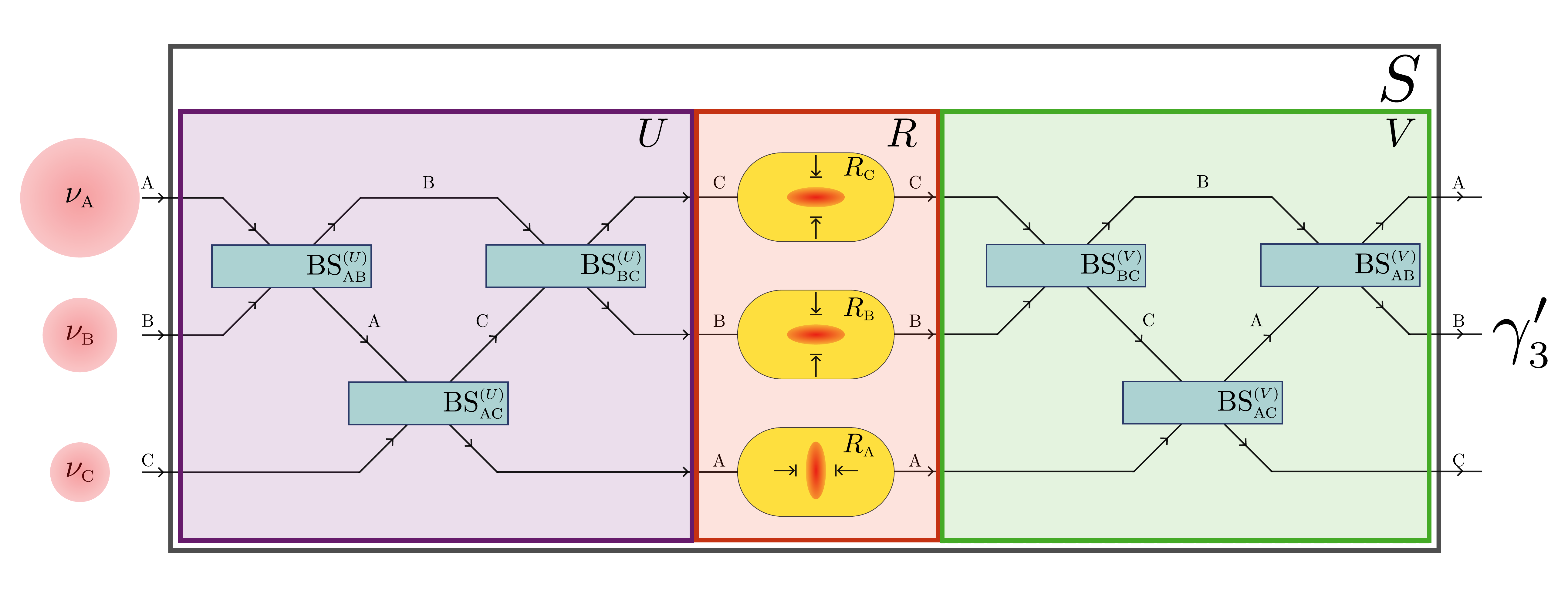}
\caption{Decomposition of symplectic transformation $S$ generating a Gaussian
state with CM $\gamma_{3}$ of three modes $A,B$ and $C$: $\nu_{j}$ -- thermal
states with mean number of thermal photons $(\nu_{j}-1)/2$, $j=A,B,C$ (red circles);
$U$ -- passive transformation consisting of beam splitters $\mathrm{BS}_{jk}^{(U)}$, $jk=AB, AC, BC$ (magenta block);
$V$ -- passive transformation consisting of beam splitters $\mathrm{BS}_{jk}^{(V)}$ (green block); $R$ -- squeezing transformation consisting of one squeezer in position quadrature, $R_{A}$,
and two squeezers in momentum quadrature, $R_{B}$ and $R_{C}$ (pink block). For rounded parameters as in Tabs.~\ref{tab:symp} and \ref{tab:bs} the circuit produces the
CM $\gamma'_{3}$, which closely approximates the CM $\gamma_{3}$, and retains its entanglement properties. See text for details.}
\label{fig2}
\end{center}
\end{figure}

In the previous section we have seen that the investigated effect is strongest in the three-mode case.
For this reason, we now derive a linear-optical scheme for preparing a Gaussian state with
the three-mode CM $\gamma_{3}$. The scheme is depicted in Fig.~\ref{fig2}.

The scheme follows from Williamson's symplectic diagonalisation of a CM
\cite{Williamson_36}, the Bloch-Messiah decomposition of a symplectic matrix \cite{Braunstein_05} and
the decomposition of an orthogonal symplectic matrix into an array of beam splitters and phase-shifters
\cite{Reck_94,vanLoock_02}. More precisely, according to Williamson's theorem \cite{Williamson_36}
for any CM $\gamma$ there is a symplectic transformation $\mathscr{S}$ which brings the CM to the normal
form,
%%%%%%%%%%%%%%%%%%%%%%%%%%%%%%%%%%%%%%%%%%%%%%%%%%%%%%%%%%%%%%%%%%%%%%%%%%%%%%%%%%%%%%%%%%%%%%%%%%%%%
\begin{equation}\label{Williamson}
\mathscr{S}\gamma\mathscr{S}^{T}=\bigoplus_{i=1}^{N}\nu_{i}\openone\equiv W,
\end{equation}
%%%%%%%%%%%%%%%%%%%%%%%%%%%%%%%%%%%%%%%%%%%%%%%%%%%%%%%%%%%%%%%%%%%%%%%%%%%%%%%%%%%%%%%%%%%%%%%%%%%%%
where $\nu_{1},\nu_{2},\ldots,\nu_{N}\geq1$ are the so called symplectic
eigenvalues of CM $\gamma$. In particular, $\nu_{1}=\nu_{2}=\ldots=\nu_{N}=1$ if the state is pure.
Consequently, making use of the symplectic transformation $S\equiv\mathscr{S}^{-1}$, one can write $\gamma = S W S^T$.
The symplectic eigenvalues are the magnitudes of the eigenvalues of the matrix $i \Omega \gamma$ \cite{Vidal_02} and
for CM $\gamma_3$ they are written in Tab. \ref{tab:symp}. The corresponding symplectic matrix $S$ can be found numerically
either using a method of Ref.~\cite{Serafini_05} or a method of Ref.~\cite{Pirandola_09}.

Making use of the Bloch-Messiah decomposition \cite{Braunstein_05} we further numerically decomposed the symplectic matrix $S$ into
passive transformations $U$ and $V$, and an active transformation $R$, as
%%%%%%%%%%%%%%%%%%%%%%%%%%%%%%%%%%%%%%%%%%%%%%%%%%%%%%%%%%%%%%%%%%%%%%%%%%%%
\begin{equation}
\label{S}
S = V R\, U.
\end{equation}
%%%%%%%%%%%%%%%%%%%%%%%%%%%%%%%%%%%%%%%%%%%%%%%%%%%%%%%%%%%%%%%%%%%%%%%%%%%
Here, $U$ and $V$ are orthogonal and symplectic transformations and
$R = R_A(s_A) \oplus R_B(s_B^{-1}) \oplus R_C(s_C^{-1})$ is the squeezing transformation,
where $R_j(s_j) = \mbox{diag}(s_j,s_j^{-1})$, $j=A,B,C$, is the diagonal matrix and the squeezing
parameters $s_{j}<1$ may be found in Tab.~\ref{tab:symp}. These transformations are highlighted by the
coloured boxes in Fig.~\ref{fig2}.

\begin{table}[h]
    \caption{Symplectic eigenvalues $\nu_j$ and the squeezing parameters $s_j$.}
\centering
\begin{tabular}{|c|c|c|c|}
\hline
$j$      & A  & B & C  \\
\hline
$\nu_{j}$   & 6.835 & 1.012 & 1.004 \\
%$\nu_{j}$   & 6.83477   & 1.01169   & 1.00379
\hline
$s_{j}$    & 0.396 & 0.851 & 0.478 \\
%$s_{j}$    & 1/2.52523& 0.850531& 0.477678\\
\hline
\end{tabular}
\label{tab:symp}
\end{table}

Next, following the method of Refs.~\cite{Reck_94,vanLoock_02} one can decompose the passive
transformations $U$ and $V$ into an array of three beam splitters as in Fig.~\ref{fig2},
%%%%%%%%%%%%%%%%%%%%%%%%%%%%%%%%%%%%%%%%%%%%%%%%%%%%%%%%%%%%%%%%%%%%%%%%%%%%%%%
\begin{eqnarray}
\label{UV}
U&=&B^{(U)}_{BC}(T_{BC}) B^{(U)}_{AC}(T_{AC}) B^{(U)}_{AB}(T_{AB}),\nonumber\\
V&=&B^{(V)}_{AB}(\tau_{AB})B^{(V)}_{AC}(\tau_{AC})B^{(V)}_{BC}(\tau_{BC}),
\end{eqnarray}
%%%%%%%%%%%%%%%%%%%%%%%%%%%%%%%%%%%%%%%%%%%%%%%%%%%%%%%%%%%%%%%%%%%%%%%%%%%%%%%%%%%%%%%%%
where the beam splitter matrices $B^{(U)}_{jk}(T_{jk})$ and $B^{(V)}_{jk}(\tau_{jk})$, $jk = AB,AC,BC$,
are given explicitly in Appendix~\ref{sec_app_beamsplitters}, and
the beam splitter transmissivities $T_{jk}$ and $\tau_{jk}$ can be found in Tab.~\ref{tab:bs}.

\begin{table}[h]
    \caption{Amplitude transmissivities $T_{jk}$ and $\tau_{jk}$.}
\centering
\begin{tabular}{|c|c|c|c|}
\hline
$jk$        & AB        & AC        & BC        \\
\hline
$T_{jk}$    & 0.555     & 0.947     & 0.492     \\
%$T_{jk}$    & 0.554681& 0.94748& 0.492407 \\
% Inverse values
%$\sqrt{1-T_{jk}^2}$    &0.835165& 0.31225& 0.871722\\
\hline
$\tau_{jk}$ & 0.716     & 0.904     & 0.657     \\
%$\tau_{jk}$ & 0.715942& 0.903823& 0.656702\\
%Inverse values
%$\sqrt{1-\tau^2_{jk}}$ &0.693974& 0.43589& 0.751266\\
\hline
\end{tabular}
\label{tab:bs}
\end{table}

Needless to say, our decomposition is numerical and thus we rounded its
parameters to three decimal places.
Consequently, the output CM $\gamma_{3}'$ slightly deviates from the original CM $\gamma_{3}$,
yet it retains all relevant entanglement properties:  it is genuinely multipartite entangled
with $\mbox{Tr}[\gamma_{3}' Z_{3}']-1 = -0.138$; and the marginals are all separable as per Tab.~\ref{tab:ppt_gamma'}.
%%%%%%%%%%%%%%%%%%%%%%%%%%%%%%%%%%%%%%%%%%%%%%%%%%%%%%%%%%%%%%%%%%%%%%%%%%%%%%%%%%%%%%%%%%%%%%%%%%%%%%%%%%%%%%%%%%%
\begin{table}[ht]
\caption{Minimal eigenvalue
${\varepsilon}_{jk}'\equiv\mathrm{min}\{\mathrm{eig}[\gamma_{3,jk}'^{(T_{j})}+i\Omega_{2}]\}$.} \centering
\begin{tabular}{| c | c | c | c |}
\hline $jk$                 & AB    & AC    & BC     \\
\hline
$\varepsilon_{jk}$   &0.005      & 0.852     & 0.010         \\
%$\varepsilon_{jk}$   &0.00527321 & 0.851925  & 0.00965682    \\
\hline
\end{tabular}
\label{tab:ppt_gamma'}
\end{table}
%%%%%%%%%%%%%%%%%%%%%%%%%%%%%%%%%%%%%%%%%%%%%%%%%%%%%%%%%%%%%%%%%%%%%%%%%%%%%%%%%%%%%%%%%%%%%%%%%%%%%%%%%%%%%%%%%%
The CM $\gamma_{3}'$ and the corresponding witness $Z_{3}'$ can be found in Appendix \ref{app_circuit_experimental}.

In the next section, we present an equivalent, yet simpler, circuit whose output
CM still retains all required properties.
%%%%%%%%%%%%%%%%%%%%%%%%%%%%%%%%%%%%%%%%%%%%%%%%%%%%%%%%%%%%%%%%%%%%%%%%%%%%%%%%%%%%%%%%%%%%%%%%%%%%%%%%%%%%%%%%%
\subsection{Simplified circuit}\label{sec_scheme_simplified}

\begin{figure}[ht]
\begin{center}
\includegraphics[width=0.48\textwidth]{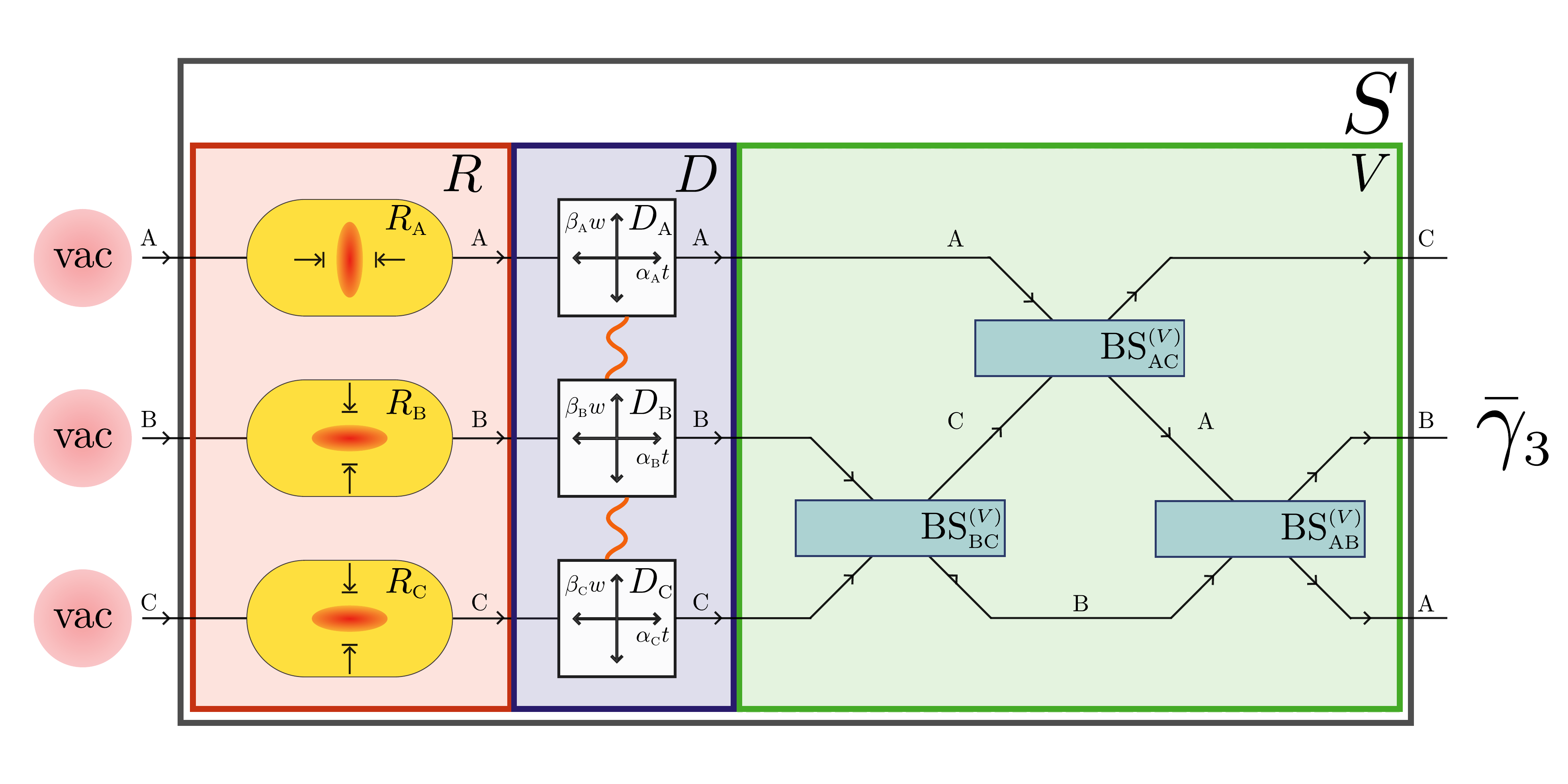}
\caption{Scheme for preparation of a Gaussian state with CM $\bar{\gamma}_{3}$ carrying genuine multipartite entanglement verifiable
from nearest-neighbour separable marginals. The input comprises of three vacuum states (red circles). The squeezing transformation $R$ (red box)
and the transformation $V$ (green box) are the same as in Fig.~\ref{fig2}. The block $D$ (gray box) contains correlated displacements
$D_{A}, D_{B}$ and $D_{C}$ (white boxes) given in Eq.~(\ref{displacements}), where the parameters $\alpha_{j}$ and $\beta_{j}$
are in Tab.~\ref{tab:displacement} and $\langle t^2 \rangle = \langle w^2 \rangle = (\nu_A-1)/2$. See text for details.}
\label{fig3}
\end{center}
\end{figure}

The scheme in Fig.~\ref{fig2} offers two simplifications which make its
experimental realization easier. First, inspecting Tab.~\ref{tab:symp}
one may see that the input states of modes $B$ and $C$ can be approximated by vacuua.
Second, the classically correlated state subject to the squeezing transformations can be
replaced by correlatively displaced squeezed vacuum states. This follows from the
fact that a thermal state at the input of mode $A$ can be prepared by the
displacements $x_{A}^{(0)}\rightarrow x_{A}^{(0)}+t$ and $p_{A}^{(0)}\rightarrow p_{A}^{(0)}+w$
of its position and momentum vacuum quadratures $x_{A}^{(0)}$ and $p_{A}^{(0)}$, respectively,
where $t$ and $w$ are uncorrelated classical zero mean Gaussian random variables with
second moments $\langle t^2 \rangle = \langle w^2 \rangle = (\nu_A - 1)/2$. As on the level of quadrature operators
the transformations $U$ and $R$ are linear, we can push the displacements through the transformations so that behind
the transformation $R$ they attain the following form:
%%%%%%%%%%%%%%%%%%%%%%%%%%%%%%%%%%%%%%%%%%%%%%%%%%%%%%%%%%%%%%%%%%%%%%%%%%%%%%%%%%%%%%%%%%%%%%%%%%%%%%%%%%%
\begin{equation}\label{displacements}
x_{j}\rightarrow x_{j}+\alpha_{j}t,\quad p_{j}\rightarrow p_{j}+\beta_{j}w,
\end{equation}
%%%%%%%%%%%%%%%%%%%%%%%%%%%%%%%%%%%%%%%%%%%%%%%%%%%%%%%%%%%%%%%%%%%%%%%%%%%%%%%%%%%%%%%%%%%%%%%%%%%%%%%%%
$j=A,B,C$, where the parameters $\alpha_{j}$ and $\beta_{j}$ after rounding read as in Tab.~\ref{tab:displacement}.
%%%%%%%%%%%%%%%%%%%%%%%%%%%%%%%%%%%%%%%%%%%%%%%%%%%%%%%%%%%%%%%%%%%%%%%%%%%%%%%%%%%%%%%%%%%%%%%%%%%%%%%
\begin{table}[h]
    \caption{Parameters $\alpha_{j}$ and $\beta_{j}$ of displacements (\ref{displacements}).}
\centering
\begin{tabular}{|c|c|c|c|}
    \hline
    $j$ & A   & B & C  \\
    \hline
    $\alpha_j$
    &   0.2
    & - 0.7
    &   1.3     \\
    %$\alpha_j$ &   0.208119 & - 0.663247 &   1.33322  \\
    \hline
    $\beta_j$
    &   1.3
    & - 0.5
    &   0.3     \\
    %$\beta_j$ &   1.32713  & - 0.479795 &   0.304208 \\
    \hline
\end{tabular}
\label{tab:displacement}
\end{table}
%%%%%%%%%%%%%%%%%%%%%%%%%%%%%%%%%%%%%%%%%%%%%%%%%%%%%%%%%%%%%%%%%%%%%%%%%%%%%%%%%%%%%%%%%%%%%%%%%%%%%%%%%%%%%%%%%
Further, the first step of the obtained scheme consists of application of a passive transformation $U$ on three
vacuum states, which is nothing but a triple of vacuum states and thus the transformation $U$ can be
omitted completely. In this way, we arrive at the simplified scheme depicted in Fig.~\ref{fig3}.

%%%%%%%%%%%%%%%%%%%%%%%%%%%%%%%%%%%%%%%%%%%%%%%%%%%%%%%%%%%%%%%%%%%%%%%%%%%%%%%%%%%%%%%%%%%%%%%%%%%%%%
%%%%%%%%%%%%%%%%%%%%%%%%%%%%%%%%%%%%%%%%%%%%%%%%%%%%%%%%%%%%%%%%%%%%%%%%%%%%%%%%%%%%%%%%%%%%%%%%%%%%%%%%

Using the squeezing parameters and transmissivities, found in the second row of
Tabs.~\ref{tab:symp} and \ref{tab:bs} respectively, as well as the displacements
in Tab.~\ref{tab:displacement}, the circuit in Fig.~\ref{fig3} produces a state
which is genuinely multipartite entangled and has all marginals separable.
Calling the CM of this state $\bar{\gamma}_3$, the optimal witness
for this CM gives $\mbox{Tr}[\bar{\gamma}_3 \bar{Z}_3] - 1 = -0.139$.  The
numerical CM $\bar{\gamma}_3$ along with the corresponding entanglement witness
may be found in Appendix \ref{app_circuit_experimental}.

The simplified scheme in Fig.~\ref{fig3} makes experimental demonstration of the
investigated effect more viable. Primarily, preparation of squeezed states at
the input is easier than implementation of squeezing operations in between beam
splitter arrays $U$ and $V$ (compare positions of pink boxes $R$ in
Figs.~\ref{fig2} and \ref{fig3}). Further, the largest amount of squeezing,
$10 \mbox{Log}_{10}[(s_A)^2] \doteq -8\, \mbox{dB}$,
%$10 \mbox{Log}_{10}[(s_A)^2] \doteq -8.04609\, \mbox{dB}$,
is well within the reach of the current technology \cite{Vahlbruch_16}, and what is more, 
one may decrease the squeezing required at the cost of decreased effect strength.
Additionally, the effect is immune to rounding of CMs and some parameters of the circuit
components, which indicates, that perfect matching of the setup parameters with the theoretical
values is not critical for its demonstration. Finally, as we have already mentioned, the output state 
tolerates the addition of a small amount of thermal noise, which is, however, of
the same size as for the comparably fragile, yet 
already demonstrated similarly complex setup \cite{DiGuglielmo_11}. The extent to which the relatively low noise 
tolerance and other imperfections are detrimental to observability of the
investigated phenomenon depends on the used experimental 
platform and will be addressed elsewhere.

%%%%%%%%%%%%%%%%%%%%%%%%%%%%%%%%%%%%%%%%%%%%%%%%%%%%%%%%%%%%%%%%%%%%%%%%%%%%%%%%%%%%%%%%%%%%%%%%%%%%%%%%%%%%
\section{Conclusions}\label{sec_discussion}

In this paper we extended the concept of genuine multipartite entanglement verifiable from separable marginals to the
domain of Gaussian states. We constructed many examples of Gaussian states possessing all two-mode marginals
separable and whose genuine multipartite entanglement can be certified solely from the set of nearest-neighbour
marginals. Each of the sets is characterized by a connected graph with no cycles, where the vertices represent the modes and the edges
the nearest-neighbour marginals. Our examples are numerical and result from an iterative search algorithm
relying on construction of a genuine multipartite witness in the space of covariance matrices. Moreover, the witness
is `blind' to correlations between modes corresponding to non-adjacent vertices in the respective graph.

Here, we gave examples for all configurations encompassing three and four modes thus complementing the study of the investigated
phenomenon in multi-qubit systems \cite{Paraschiv_17}. The three-mode state found by us
exhibits the strongest form of the property compared to the four-mode cases and therefore we also proposed a
scheme for preparation of the state, which consists of three quadrature squeezers sandwiched between two triples of
phase-free beam splitters. Further, we replaced the original scheme by a simpler scheme, which still produces the desired effect,
but requires only interference of three squeezed states subjected to correlated displacements
on three beam splitters. The squeezing used in the setup is well within the reach of the
current technology. Additionally, all relevant properties of the output state remain preserved after
contamination by a small amount of thermal noise which gives us a hope that the investigated property
of genuine multipartite entanglement could be observed. A successful realization of the proposed setup
would mean extension of the experimental analysis of the phenomenon of emergent genuine multipartite
entanglement \cite{Miklin_16} from qubits and the scenario when all bipartite marginals are known
\cite{Micuda_19}, to the realm of Gaussian states and more generic situation when only some bipartite
marginals are known.

The impact of the presented results is twofold. On one hand, they point at an alternative approach
towards experimental investigation of the remarkable concept of genuine multipartite entanglement
verifiable from incomplete sets of separable marginals. On the other hand, they also stimulate theoretical questions
concerning the existence of a Gaussian classical analog of the quantum marginal problem \cite{Leskovjanova_20} or the extendibility of the
entanglement marginal problem \cite{Navascues_20} to Gaussian case. On a more general level, our results
contribute to the development of methods of detection of global properties of multipartite quantum systems
from partial information.

%%%%%%%%%%%%%%%%%%%%%%%%%%%%%%%%%%%%%%%%%%%%%%%%%%%%%%%%%%%%%%%%%%%%%%%%%%%%%%%%%%%%%%%%%%%%%%%%%%%%%%%%%%%%%%%%%%%%%%%%%%%%%%%%%%%%%%%%%%%%%%%

\acknowledgments
O.L. and J.P. acknowledge support from IGA-PrF-2020-009 and IGA-PrF-2021-006.
J.P. also acknowledges project GA18-21285S of the Grant Agency of Czech
Republic (GAČR).
V.N. and N.K. have been supported by the Scottish Universities Physics
Alliance (SUPA) and by the Engineering and Physical Sciences Research Council
(EPSRC). N.K. was supported by the EU Flagship on Quantum Technologies, project
PhoG (820365).

\section*{\noindent Competing interests}
\vspace{-1.3em}
\noindent The authors declare that there are no competing interests.
\vspace{-2em}

\section*{Author contribution}
\vspace{-1.3em}
\noindent L.M. conceived the theory,
J.P., V.N. and O.L. contributed to writing of the code,
V.N., O.L., J.P. and L.M. carried out calculations,
V.N., L.M. and N.K. wrote the manuscript,
N.K. and L.M. supervised the project,
all authors discussed the paper.
\vspace{-2em}

\section*{Data availability}
\vspace{-1.3em}
\noindent The data generated or analysed during the current study are available
from the corresponding author upon reasonable request.
\vspace{-1.5em}

\appendix
\section{Block-diagonal matrices in SDP (\ref{bisepdual})}\label{sec_app_I}

In this section we give an explicit form of matrices $X_{j}^{\mathrm{bd},\pi(k)}$ appearing in SDP (\ref{bisepdual}) for $N=3$ and $N=4$.

%%%%%%%%%%%%%%%%%%%%%%%%%%%%%%%%%%%%%%%%%%%%%%%%%%%%%%%%%%%%%%%%%%%%%%%%%%%%%%%%%%%%%%%%%%
\subsection{$\mathbf{N=3}$}\label{subsec_app_three-mode_Xjbd}

For $N=3$ we have altogether $K=3$ bipartitions $\pi(1)=A|BC, \pi(2)=B|AC$ and $\pi(3)=C|AB$, where we have omitted the curly brackets from the lists of elements
of the sets $\mathcal{M}_{\mathcal{J}_{k}}$ and $\bar{\mathcal{M}}_{\mathcal{J}_{k}}$ for brevity. The first equality in SDP (\ref{bisepdual}) imposes constraints
on certain elements of real parts of $6\times 6$ Hermitian matrices $X_{j}$, $j=1,2,3,4$, which are embodied into matrices, $X_{j}^{\mathrm{bd},\pi(k)}$, given explicitly as
%%%%%%%%%%%%%%%%%%%%%%%%%%%%%%%%%%%%%%%%%%%%%%%%%%%%%%%%%%%%%%%%%%%%%%%%%%%%%%%%%%%%
\begin{equation*}\label{Xjbdpi1}
X_{j}^{\mathrm{bd},\pi(1)}=\left(\begin{array}{ccc}
(X_{j})_{11} & \mathbb{O} & \mathbb{O}  \\
\mathbb{O} & (X_{j})_{22} & (X_{j})_{23}  \\
\mathbb{O} & (X_{j})_{23}^{\dag} & (X_{j})_{33}  \\
\end{array}\right),
\end{equation*}
%%%%%%%%%%%%%%%%%%%%%%%%%%%%%%%%%%%%%%%%%%%%%%%%%%%%%%%%%%%%%%%%%%%%%%%%%%%%%%%%%%%%%%%%%%%%%%%%%%%%%

%%%%%%%%%%%%%%%%%%%%%%%%%%%%%%%%%%%%%%%%%%%%%%%%%%%%%%%%%%%%%%%%%%%%%%%%%%%%%%%%%%%%
\begin{equation*}\label{Xjbdpi2}
X_{j}^{\mathrm{bd},\pi(2)}=\left(\begin{array}{ccc}
(X_{j})_{11} & \mathbb{O} & (X_{j})_{13}  \\
\mathbb{O} & (X_{j})_{22} & \mathbb{O}  \\
(X_{j})_{13}^{\dag} & \mathbb{O} & (X_{j})_{33}  \\
\end{array}\right),
\end{equation*}
%%%%%%%%%%%%%%%%%%%%%%%%%%%%%%%%%%%%%%%%%%%%%%%%%%%%%%%%%%%%%%%%%%%%%%%%%%%%%%%%%%%%%%%%%%%%%%%%%%%%%

%%%%%%%%%%%%%%%%%%%%%%%%%%%%%%%%%%%%%%%%%%%%%%%%%%%%%%%%%%%%%%%%%%%%%%%%%%%%%%%%%%%%
\begin{equation*}\label{Xjpi3}
X_{j}^{\mathrm{bd},\pi(3)}=\left(\begin{array}{ccc}
(X_{j})_{11} & (X_{j})_{12} & \mathbb{O}  \\
(X_{j})_{12}^{\dag} & (X_{j})_{22} & \mathbb{O}  \\
\mathbb{O} & \mathbb{O} & (X_{j})_{33}  \\
\end{array}\right).
\end{equation*}
%%%%%%%%%%%%%%%%%%%%%%%%%%%%%%%%%%%%%%%%%%%%%%%%%%%%%%%%%%%%%%%%%%%%%%%%%%%%%%%%%%%%%%%%%%%%%%%%%%%%%

%%%%%%%%%%%%%%%%%%%%%%%%%%%%%%%%%%%%%%%%%%%%%%%%%%%%%%%%%%%%%%%%%%%%%%%%%%%%%%%%%%%%%%%%%%
\subsection{$\mathbf{N=4}$}\label{subsesec_app_four-mode_Xjbd}

For $N=4$ there are $K=7$ bipartitions $\pi(1)=A|BCD,\pi(2)=B|ACD,\pi(3)=C|ABD,\pi(4)=D|ABC, \pi(5)=AB|CD,\pi(6)=AC|BD$ and $\pi(7)=AD|BC$.
The matrices $X_{j}^{\mathrm{bd},\pi(k)}$, $k=1,\ldots,7$, obtained by projection of the matrices $X_{j}$ onto the block-diagonal form corresponding to bipartiton $\pi(k)$ read explicitly as
%%%%%%%%%%%%%%%%%%%%%%%%%%%%%%%%%%%%%%%%%%%%%%%%%%%%%%%%%%%%%%%%%%%%%%%%%%%%%%%%%%%%
%%%%%%%%%%%%%%%%%%%%%%%%%%%%%%%%%%%%%%%%%%%%%%%%%%%%%%%%%%%%%%%%%%%%%%%%%%%%%%%%%%%%
\begin{equation*}\label{Xjbdpi1N4}
X_{j}^{\mathrm{bd},\pi(1)}=\left(\begin{array}{cccc}
(X_{j})_{11} & \mathbb{O} & \mathbb{O} & \mathbb{O} \\
\mathbb{O} & (X_{j})_{22} & (X_{j})_{23}  & (X_{j})_{24}  \\
\mathbb{O} & (X_{j})_{23}^{\dag} & (X_{j})_{33} & (X_{j})_{34} \\
\mathbb{O} & (X_{j})_{24}^{\dag} & (X_{j})_{34}^{\dag} & (X_{j})_{44} \\
\end{array}\right),
\end{equation*}
%%%%%%%%%%%%%%%%%%%%%%%%%%%%%%%%%%%%%%%%%%%%%%%%%%%%%%%%%%%%%%%%%%%%%%%%%%%%%%%%%%%%%%%%%%%%%%%%%%%%%

%%%%%%%%%%%%%%%%%%%%%%%%%%%%%%%%%%%%%%%%%%%%%%%%%%%%%%%%%%%%%%%%%%%%%%%%%%%%%%%%%%%%%%%%%%%%%%%%%%%%%
\begin{equation*}\label{Xjbdpi2N4}
X_{j}^{\mathrm{bd},\pi(2)}=\left(\begin{array}{cccc}
(X_{j})_{11} & \mathbb{O} & (X_{j})_{13}  & (X_{j})_{14} \\
\mathbb{O} & (X_{j})_{22} & \mathbb{O}  & \mathbb{O}   \\
(X_{j})_{13}^{\dag} & \mathbb{O} & (X_{j})_{33} & (X_{j})_{34} \\
(X_{j})_{14}^{\dag} &  \mathbb{O}  & (X_{j})_{34}^{\dag} & (X_{j})_{44} \\
\end{array}\right),
\end{equation*}
%%%%%%%%%%%%%%%%%%%%%%%%%%%%%%%%%%%%%%%%%%%%%%%%%%%%%%%%%%%%%%%%%%%%%%%%%%%%%%%%%%%%%%%%%%%%%%%%%%%%%

%%%%%%%%%%%%%%%%%%%%%%%%%%%%%%%%%%%%%%%%%%%%%%%%%%%%%%%%%%%%%%%%%%%%%%%%%%%%%%%%%%%%%%%%%%%%%%%%%%%%%
\begin{equation*}\label{Xjbdpi3N4}
X_{j}^{\mathrm{bd},\pi(3)}=\left(\begin{array}{cccc}
(X_{j})_{11} & (X_{j})_{12}  & \mathbb{O} & (X_{j})_{14} \\
(X_{j})_{12}^{\dag} & (X_{j})_{22} & \mathbb{O}  &  (X_{j})_{24}   \\
\mathbb{O} & \mathbb{O}  & (X_{j})_{33} & \mathbb{O} \\
(X_{j})_{14}^{\dag} & (X_{j})_{24}^{\dag} & \mathbb{O} & (X_{j})_{44} \\
\end{array}\right),
\end{equation*}
%%%%%%%%%%%%%%%%%%%%%%%%%%%%%%%%%%%%%%%%%%%%%%%%%%%%%%%%%%%%%%%%%%%%%%%%%%%%%%%%%%%%%%%%%%%%%%%%%%%%%

%%%%%%%%%%%%%%%%%%%%%%%%%%%%%%%%%%%%%%%%%%%%%%%%%%%%%%%%%%%%%%%%%%%%%%%%%%%%%%%%%%%%%%%%%%%%%%%%%%%%%
\begin{equation*}\label{Xjbdpi4N4}
X_{j}^{\mathrm{bd},\pi(4)}=\left(\begin{array}{cccc}
(X_{j})_{11} & (X_{j})_{12} & (X_{j})_{13} & \mathbb{O} \\
(X_{j})_{12}^{\dag} & (X_{j})_{22} & (X_{j})_{23} & \mathbb{O} \\
(X_{j})_{13}^{\dag} & (X_{j})_{23}^{\dag}  & (X_{j})_{33} & \mathbb{O} \\
\mathbb{O} & \mathbb{O} & \mathbb{O} & (X_{j})_{44} \\
\end{array}\right),
\end{equation*}
%%%%%%%%%%%%%%%%%%%%%%%%%%%%%%%%%%%%%%%%%%%%%%%%%%%%%%%%%%%%%%%%%%%%%%%%%%%%%%%%%%%%%%%%%%%%%%%%%%%%%

%%%%%%%%%%%%%%%%%%%%%%%%%%%%%%%%%%%%%%%%%%%%%%%%%%%%%%%%%%%%%%%%%%%%%%%%%%%%%%%%%%%%%%%%%%%%%%%%%%%%%
\begin{equation*}\label{Xjbdpi5N4}
X_{j}^{\mathrm{bd},\pi(5)}=\left(\begin{array}{cccc}
(X_{j})_{11} & (X_{j})_{12}  & \mathbb{O}  & \mathbb{O}  \\
(X_{j})_{12}^{\dag} & (X_{j})_{22} & \mathbb{O}  &  \mathbb{O}    \\
\mathbb{O} & \mathbb{O}  & (X_{j})_{33} & (X_{j})_{34} \\
\mathbb{O}  & \mathbb{O}  & (X_{j})_{34}^{\dag} & (X_{j})_{44} \\
\end{array}\right),
\end{equation*}
%%%%%%%%%%%%%%%%%%%%%%%%%%%%%%%%%%%%%%%%%%%%%%%%%%%%%%%%%%%%%%%%%%%%%%%%%%%%%%%%%%%%%%%%%%%%%%%%%%%%%

%%%%%%%%%%%%%%%%%%%%%%%%%%%%%%%%%%%%%%%%%%%%%%%%%%%%%%%%%%%%%%%%%%%%%%%%%%%%%%%%%%%%%%%%%%%%%%%%%%%%%
\begin{equation*}\label{Xjbdpi6N4}
X_{j}^{\mathrm{bd},\pi(6)}=\left(\begin{array}{cccc}
(X_{j})_{11} & \mathbb{O} & (X_{j})_{13}  & \mathbb{O}  \\
\mathbb{O} & (X_{j})_{22} & \mathbb{O}  & (X_{j})_{24}    \\
(X_{j})_{13}^{\dag} & \mathbb{O}  & (X_{j})_{33} & \mathbb{O} \\
\mathbb{O}  & (X_{j})_{24}^{\dag} & \mathbb{O}  & (X_{j})_{44} \\
\end{array}\right),
\end{equation*}
%%%%%%%%%%%%%%%%%%%%%%%%%%%%%%%%%%%%%%%%%%%%%%%%%%%%%%%%%%%%%%%%%%%%%%%%%%%%%%%%%%%%%%%%%%%%%%%%%%%%%

%%%%%%%%%%%%%%%%%%%%%%%%%%%%%%%%%%%%%%%%%%%%%%%%%%%%%%%%%%%%%%%%%%%%%%%%%%%%%%%%%%%%%%%%%%%%%%%%%%%%%
\begin{equation*}\label{Xjbdpi7N4}
X_{j}^{\mathrm{bd},\pi(7)}=\left(\begin{array}{cccc}
(X_{j})_{11} & \mathbb{O} & \mathbb{O}  & (X_{j})_{14} \\
\mathbb{O} & (X_{j})_{22} & (X_{j})_{23}   & \mathbb{O}   \\
\mathbb{O} & (X_{j})_{23}^{\dag} & (X_{j})_{33} & \mathbb{O} \\
(X_{j})_{14}^{\dag} &  \mathbb{O}  & \mathbb{O} & (X_{j})_{44} \\
\end{array}\right).
\end{equation*}
%%%%%%%%%%%%%%%%%%%%%%%%%%%%%%%%%%%%%%%%%%%%%%%%%%%%%%%%%%%%%%%%%%%%%%%%%%%%%%%%%%%%%%%%%%%%%%%%%%%%%

\section{Four-mode numerical examples}\label{sec_app_II}

We give explicit form of numeric witnesses for
the four-mode CMs $\gamma_{4}^{(1)}$ and $\gamma_{4}^{(2)}$ detecting genuine
multipartite entanglement from minimal sets of two-mode marginal CMs
characterized by the linear tree and the `t'-shaped tree in Figs.~\ref{fig1} b) and c), respectively.

\subsection{Linear tree}

The witness which detects the genuine multipartite entanglement of CM $\gamma_{4}^{(1)}$ without
accessing correlations between pairs of modes $(A,C), (A,D)$ and $(B,D)$ is
\begin{widetext}
\begin{eqnarray*}\label{appWitnessLinear}
    Z_{4}^{(1)}=10^{-2} \cdot \left(\begin{array}{cccccccc}
    2.70  &      0 & -1.12  &      0 &      0 &      0 &      0 &      0\\
        0 & 33.29  &      0 &-28.67  &      0 &      0 &      0 &      0\\
   -1.12  &      0 &  6.86  &      0 &  6.30  &      0 &      0 &      0\\
        0 &-28.67  &      0 & 29.50  &      0 & -5.46  &      0 &      0\\
        0 &      0 &  6.30  &      0 & 74.73  &      0 & 33.42  &      0\\
        0 &      0 &      0 & -5.46  &      0 &  7.37  &      0 &  2.18 \\
        0 &      0 &      0 &      0 & 33.42  &      0 & 16.30  &      0\\
        0 &      0 &      0 &      0 &      0 &  2.18  &      0 &  4.11
\end{array}\right).
\end{eqnarray*}
\end{widetext}

\subsection{`t'-shaped tree}

The witness detecting genuine multipartite entanglement of CM $\gamma_{4}^{(2)}$, which is `blind' with respect to correlations between the pairs of modes $(A,C), (A,D), (C,D)$, reads as
%%%%%%%%%%%%%%%%%%%%%%%%%%%%%%%%%%%%%%%%%%%%%%%%%%%%%%%%%%%%%%%%%%%%%%%%%%%%%%%%%%%%%%%%%%%%%%
\begin{widetext}
\begin{eqnarray*}\label{witTshape}
    Z_{4}^{(2)}= 10^{-2} \cdot
    \left(\begin{array}{cccccccc}
    1.984  &       0 & -0.815 &        0 &       0&        0 &       0 &       0\\
         0 & 76.150  &       0& -26.031  &       0&        0 &       0 &       0\\
   -0.815  &       0 & 37.883 &        0 & -1.525 &        0 & 19.701  &       0\\
         0 &-26.031  &       0&  18.014  &       0& -22.092  &       0 & -0.760 \\
         0 &       0 & -1.525 &        0 &  2.895 &        0 &       0 &       0\\
         0 &       0 &       0& -22.092  &       0&  54.640  &       0 &       0\\
         0 &       0 & 19.701 &        0 &       0&        0 & 10.563  &       0\\
         0 &       0 &       0&  -0.760  &       0&        0 &       0 &  3.149
\end{array}\right).
\end{eqnarray*}
\end{widetext}
%%%%%%%%%%%%%%%%%%%%%%%%%%%%%%%%%%%%%%%%%%%%%%%%%%%%%%%%%%%%%%%%%%%%%%%%%%%%%%%%%%%%%%%%%%%%%%%%%%

\section{Beam splitter transformations}\label{sec_app_beamsplitters}

In this section we give explicit form of beam splitter matrices appearing in Eq.~(\ref{UV}) of the main text,
%%%%%%%%%%%%%%%%%%%%%%%%%%%%%%%%%%%%%%%%%%%%%%%%%%%%%%%%%%%%%%%%%%%%%%%%%%%%%%%%%%%%%%%%%%%%%%%%%%%%%%%%%%%
\begin{equation*}
    \label{bb:U_AB}
    B^{(U)}_{AB}(T_{AB})
    =
    \left(
        \begin{matrix}
            T_{AB}\openone & R_{AB}\openone & \mathbb{O}\\
            R_{AB}\openone & -T_{AB}\openone & \mathbb{O} \\
            \mathbb{O} & \mathbb{O} & -\openone
        \end{matrix}
    \right),
\end{equation*}
%%%%%%%%%%%%%%%%%%%%%%%%%%%%%%%%%%%%%%%%%%%%%%%%%%%%%%%%%%%%%%%%%%%%%%%%%%%%%%%%%%%%%%%%%%%%%%%%%%%%%%%%%%
\begin{equation*}
    \label{bb:U_AC}
    B^{(U)}_{AC}(T_{AC})
    =
    \left(
        \begin{matrix}
            T_{AC}\openone & \mathbb{O} & R_{AC}\openone \\
            \mathbb{O}   & \openone & \mathbb{O}   \\
            R_{AC}\openone & \mathbb{O} & -T_{AC}\openone
        \end{matrix}
    \right),
\end{equation*}
%%%%%%%%%%%%%%%%%%%%%%%%%%%%%%%%%%%%%%%%%%%%%%%%%%%%%%%%%%%%%%%%%%%%%%%%%%%%%%%%%%%%%%%%%%%%%%%%%%%%%%%%%%
\begin{equation*}
    \label{bb:U_BC}
    B^{(U)}_{BC}(T_{BC})
    =
    \left(
        \begin{matrix}
            \openone   & \mathbb{O}                   & \mathbb{O}                    \\
            \mathbb{O}   & -T_{BC}\openone  & -R_{BC}\openone               \\
            \mathbb{O}   & R_{BC}\openone   & -T_{BC}\openone
        \end{matrix}
    \right),
\end{equation*}
%%%%%%%%%%%%%%%%%%%%%%%%%%%%%%%%%%%%%%%%%%%%%%%%%%%%%%%%%%%%%%%%%%%%%%%%%%%%%%%%%%%%%%%%%%%%%%%%%%%%%%%%%%
\begin{equation*}
    \label{bb:V_AB}
    B^{(V)}_{AB}(\tau_{AB})
    =
    \left(
        \begin{matrix}
            \tau_{AB}\openone & \rho_{AB}\openone & \mathbb{O}\\
            \rho_{AB}\openone & -\tau_{AB}\openone & \mathbb{O} \\
            \mathbb{O} & \mathbb{O} & \openone
        \end{matrix}
    \right),
\end{equation*}
%%%%%%%%%%%%%%%%%%%%%%%%%%%%%%%%%%%%%%%%%%%%%%%%%%%%%%%%%%%%%%%%%%%%%%%%%%%%%%%%%%%%%%%%%%%%%%%%%%%%%%%%%%
\begin{equation*}
    \label{bb:V_AC}
    B^{(V)}_{AC}(\tau_{AC})
    =
    \left(
        \begin{matrix}
            -\tau_{AC}\openone & \mathbb{O} & \rho_{AC}\openone                \\
            \mathbb{O}                   & \openone & \mathbb{O}                     \\
            -\rho_{AC}\openone              & \mathbb{O} & -\tau_{AC}\openone
        \end{matrix}
    \right),
\end{equation*}
%%%%%%%%%%%%%%%%%%%%%%%%%%%%%%%%%%%%%%%%%%%%%%%%%%%%%%%%%%%%%%%%%%%%%%%%%%%%%%%%%%%%%%%%%%%%%%%%%%%%%%%%%%
\begin{equation*}
    \label{bb:V_BC}
    B^{(V)}_{BC}(\tau_{BC})
    =
    \left(
        \begin{matrix}
            \openone   & \mathbb{O}                   & \mathbb{O}                    \\
            \mathbb{O}   & \tau_{BC}\openone & \rho_{BC}\openone               \\
            \mathbb{O}   & \rho_{BC}\openone              & -\tau_{BC}\openone
        \end{matrix}
    \right),
\end{equation*}
%%%%%%%%%%%%%%%%%%%%%%%%%%%%%%%%%%%%%%%%%%%%%%%%%%%%%%%%%%%%%%%%%%%%%%%%%%%%%%%%%%%%%%%%%%%%%%%%%%%%%%%%%%%

where the transmissivities $T_{jk}$ and $\tau_{jk}$ are given in Tab.~\ref{tab:bs} of the main text, while
$R_{jk}=\sqrt{1 - T_{jk}^2}$ and $\rho_{jk}=\sqrt{1 - \tau_{jk}^2}$ are the corresponding reflectivities.
%%%%%%%%%%%%%%%%%%%%%%%%%%%%%%%%%%%%%%%%%%%%%%%%%%%%%%%%%%%%%%%%%%%%%%%%%%%%%%%%%%%%%%%%%%%%%%%%%%%%%%%%%%%

%%%%%%%%%%%%%%%%%%%%%%%%%%%%%%%%%%%%%%%%%%%%%%%%%%%%%%%%%%%%%%%%%%%%%%%%%%%%
\section{Circuit output covariance matrices}
\label{app_circuit_experimental}

In this section we present output CMs, witnesses and relevant eigenvalues of linear-optical circuits in Figs.~~\ref{fig2} and \ref{fig3}.

\subsection{Circuit in Fig.~\ref{fig2}}

First, we present the results for the scheme in Fig.~\ref{fig2} with parameters given in Tabs.~\ref{tab:symp} and \ref{tab:bs} of the main text.
In this case the output CM, rounded to two decimal places, is given by
%%%%%%%%%%%%%%%%%%%%%%%%%%%%%%%%%%%%%%%%%%%%%%%%%%%%%%%%%%%%%%%%%%%%%%%%%%%%%%%%%%%%%%%%%%%%%%%%%%%%%%%%%%%%%%%%%%%%%%%%%%%%%%%%%%%%%%%%%%%
\begin{equation*}
    \gamma'_3 =
\left(
    \begin{array}{cccccc}
         1.34 & 0  & -0.35 & 0  & -0.82 & 0  \\
         0  & 10.01 & 0  & 8.45 & 0  & 1.86 \\
         -0.35 & 0  & 7.78 & 0  & -8.03 & 0  \\
         0  & 8.45 & 0  & 7.92 & 0  & 2.08 \\
         -0.82 & 0  & -8.03 & 0  & 9.99 & 0  \\
         0  & 1.86 & 0  & 2.08 & 0  & 1.62 \\
    \end{array}
\right).
\end{equation*}
%%%%%%%%%%%%%%%%%%%%%%%%%%%%%%%%%%%%%%%%%%%%%%%%%%%%%%%%%%%%%%%%%%%%%%%%%%%%%%%%%%%%%%%%%%%%%%%%%%%%%%%%%%%%%%%%%%%%%%%%%%%%%%%%%%%%%%%

The corresponding witness then reads as
\begin{equation*}
    Z'_3 \!\! = \!\! 10^{-2} \!\!
\left(\!\!\!\!
    \begin{array}{cccccc}
         6.86 & 0  & -0.45 & 0  & 0  & 0  \\
         0  & 34.11 & 0  & -39.31 & 0  & 0  \\
         -0.45 & 0  & 25.04 & 0  & 20.87 & 0  \\
         0  & -39.31 & 0  & 45.92 & 0  & -2.05 \\
         0  & 0  & 20.87 & 0  & 17.43 & 0  \\
         0  & 0  & 0  & -2.05 & 0  & 6.62 \\
    \end{array}
\!\!\!\!\right)
\end{equation*}
and it gives $\mbox{Tr}[\gamma_{3}' Z_{3}'] - 1 = -0.138$.
%$\mbox{Tr}[\gamma_{3}' Z'] - 1 = -0.137591$

Further, the marginals of the CMs are all separable as can be seen in
Tab.~\ref{tab:ppt_3mode'_app}.

\begin{table}[ht]
\caption{Minimal eigenvalue
${\varepsilon}_{jk}'\equiv\mathrm{min}\{\mathrm{eig}[\gamma_{3,jk}'^{(T_{j})}+i\Omega_{2}]\}$.} \centering
\begin{tabular}{| c | c | c | c |}
\hline $jk$                 & AB    & AC    & BC     \\
\hline
$\varepsilon_{jk}'$   &0.005      & 0.852     & 0.010         \\
%$\varepsilon_{jk}$   &0.00527321 & 0.851925  & 0.00965682    \\
\hline
\end{tabular}
\label{tab:ppt_3mode'_app}
\end{table}

\subsection{Circuit in Fig.~\ref{fig3}}
\label{app:simplified_circuit}

In the last section we derive and analyze entanglement properties of the CM
$\bar{\gamma}_{3}$ at the output of the circuit in Fig.~\ref{fig3}.

Initially, vacuum modes $A, B$ and $C$ enter quadrature squeezers with squeezing parameters given in the
second row of Tab.~\ref{tab:symp}. Next, they are subject to displacements
%%%%%%%%%%%%%%%%%%%%%%%%%%%%%%%%%%%%%%%%%%%%%%%%%%%%%%%%%%%%%%%%%%%%%%%%%%%%%%%%%%%%%%%%%%%%%%%%%%%%%%%%%%%
\begin{equation}\label{displacements_app}
x_{j}\rightarrow x_{j}+\alpha_{j}t,\quad p_{j}\rightarrow p_{j}+\beta_{j}w,
\end{equation}
%%%%%%%%%%%%%%%%%%%%%%%%%%%%%%%%%%%%%%%%%%%%%%%%%%%%%%%%%%%%%%%%%%%%%%%%%%%%%%%%%%%%%%%%%%%%%%%%%%%%%%%%%
where $t$ and $w$ are zero mean Gaussian random variables with second moments $\langle t^2
\rangle = \langle w^2 \rangle = ({\nu}_A - 1)/2$ and where the parameters
$\alpha_{j}$ and $\beta_{j}$ are given in Tab.~\ref{tab:displacement}. Finally,
the three modes interfere on an array of three beam splitters described by the
matrix $V$ in Eq.~(\ref{UV}). At the output of the circuit one gets the following
CM:
\begin{eqnarray*}
\bar{\gamma}_{3}=
\left(
    \begin{array}{cccccc}
         1.39 & 0  & -0.21 & 0  & -1.05 & 0  \\
         0  & 9.95 & 0  & 8.26 & 0  & 1.7 \\
         -0.21 & 0  & 7.36 & 0  & -7.83 & 0  \\
         0  & 8.26 & 0  & 7.63 & 0  & 1.94 \\
         -1.05 & 0  & -7.83 & 0  & 10.12 & 0  \\
         0  & 1.7 & 0  & 1.94 & 0  & 1.59 \\
    \end{array}
\right).
\end{eqnarray*}

The optimal witness, which gives
$\mbox{Tr}[\bar{\gamma}_3 \bar{Z}_3] - 1 = -0.139,$
%$\mbox{Tr}[\bar{gamma}_3 \bar{Z}] - 1 = -0.139075$
is given by
\begin{eqnarray*}\label{gammaTshapeApp}
    \bar{Z}_{3} \!\! = \!\! 10^{-2} \!\!
    \left( \!\!\!\!
        \begin{array}{cccccc}
             5.87 & 0  & -0.54 & 0  & 0  & 0  \\
             0  & 33.71 & 0  & -39.6 & 0  & 0  \\
             -0.54 & 0  & 26.22 & 0  & 21.01 & 0  \\
             0  & -39.6 & 0  & 47.1 & 0  & -1.87 \\
             0  & 0  & 21.01 & 0  & 16.86 & 0  \\
             0  & 0  & 0  & -1.87 & 0  & 6.17 \\
        \end{array}
    \!\!\!\! \right).
\end{eqnarray*}

All marginals are separable as evidenced by Tab.~\ref{tab:ppt_simplified}.

\begin{table}[ht]
\caption{Minimal eigenvalue
$\bar{\varepsilon}_{jk}\equiv\mathrm{min}\{\mathrm{eig}[\bar{\gamma}_{3,jk}^{(T_{j})}+i\Omega_{2}]\}$.} \centering
\begin{tabular}{| c | c | c | c |}
\hline $jk$         & AB    & AC    & BC    \\
\hline
$\bar{\varepsilon}_{jk}$   & 0.027     & 0.862     & 0.037     \\
%$\bar{\varepsilon}_{jk}$   & 0.0273197 & 0.861955  & 0.0367514 \\
\hline
\end{tabular}
\label{tab:ppt_simplified}
\end{table}
%%%%%%%%%%%%%%%%%%%%%%%%%%%%%%%%%%%%%%%%%%%%%%%%%%%%%%%%%%%%%%%%%%%%%%%%%%%%%%%%%%%%%%%%%%
%%%%%%%%%%%%%%%%%%%%%%%%%%%%%%%%%%%%%%%%%%%%%%%%%%%%%%%%%%%%%%%%%%%%%%%%%%%%%%%%%%%%%%%%%%

\end{document}